\documentclass[10pt]{spieman}  
\usepackage{amsmath,amsfonts,amssymb}
\usepackage{graphicx}
\usepackage{setspace}
\usepackage{tocloft}
\title{Direct measurement of the Kepler Space Telescope CCD's intra-pixel response function}

\author[a,b,*]{Dmitry Vorobiev}
\author[b]{Alexis Irwin}
\author[b]{Zoran Ninkov}
\author[b]{Kevan Donlon}
\author[c]{Douglas Caldwell}
\author[d]{Stefan Mochnacki}
\affil[a]{Laboratory for Atmospheric and Space Physics, University of Colorado, Boulder, Colorado, United States}
\affil[b]{Center for Imaging Science, Rochester Institute of Technology, Rochester, New York, United States}
\affil[c]{NASA Ames Research Center, Mountain View, California, United States}
\affil[d]{Department of Astronomy and Astrophysics, University of Toronto, Toronto, Ontario, Canada}

\cftpagenumbersoff{figure}
\cftpagenumbersoff{table}

\begin{document}
\twocolumn[
  \begin{@twocolumnfalse}
    \maketitle
    \begin{center} \textit{Accepted to Journal of Astronomical Telescopes, Instruments, and Systems, Special Section ``Detectors for Astronomy and Cosmology'' (September 2019).}
    \end{center}
    \begin{abstract}
Space missions designed for high precision photometric monitoring of stars often under-sample the point-spread function, with much of the light landing within a single pixel. Missions like MOST, \textit{Kepler}, BRITE, and TESS, do this to avoid uncertainties due to pixel-to-pixel response nonuniformity. This approach has worked remarkably well. However, individual pixels also exhibit response nonuniformity. Typically, pixels are most sensitive near their centers and less sensitive near the edges, with a difference in response of as much as 50\%. The exact shape of this fall-off, and its dependence on the wavelength of light, is the intra-pixel response function (IPRF). A direct measurement of the IPRF can be used to improve the photometric uncertainties, leading to improved photometry and astrometry of under-sampled systems. Using the spot-scan technique, we measured the IPRF of a flight spare e2v CCD90 imaging sensor, which is used in the \textit{Kepler} focal plane. Our spot scanner generates spots with a full-width at half-maximum of $\lesssim$3 microns across the range of 400 nm - 850 nm. We find that Kepler's CCD shows similar IPRF behavior to other back-illuminated devices, with a decrease in responsivity near the edges of a pixel by $\sim$50\%. The IPRF also depends on wavelength, exhibiting a large amount of diffusion at shorter wavelengths and becoming much more defined by the gate structure in the near-IR. This method can also be used to measure the IPRF of the CCDs used for TESS, which borrows much from the \textit{Kepler} mission. 
    \end{abstract}
    

{\noindent \footnotesize\textbf{*}Dmitry Vorobiev,  \linkable{dmitry.vorobiev@lasp.colorado.edu} }
\vspace{0.5 in}
  \end{@twocolumnfalse}
  ]

\begin{spacing}{1}   

\section{Introduction}
\label{sec:intro}  
The \textit{Kepler} mission was designed to obtain very precise photometry over a wide field to detect extrasolar planets, especially Earth-like ones, as they transit their parent stars\cite{Borucki2010}. It has been spectacularly successful, finding 2,335 confirmed exoplanets, 2,165 eclipsing binary stars, with 2,424 exoplanet candidates still to be confirmed (as of January 27, 2019). Data collection for the original \textit{Kepler} mission ended in May 2013, but the telescope continued scientific operation for 4 more years as the \textit{K2} mission, until October, 2018. Efforts are under way to develop a high quality public archive for observations acquired by \textit{Kepler} and \textit{K2}.

\begin{figure*}
    \centering
    \includegraphics[width=\textwidth]{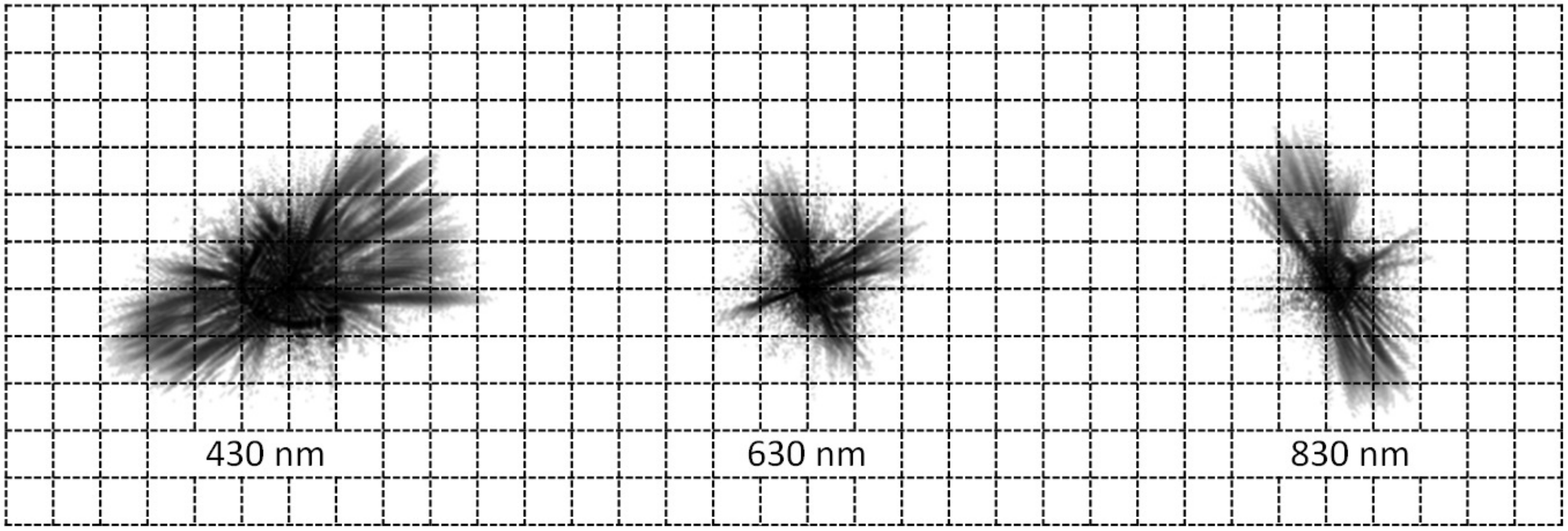}
    \caption{Synthetic monochromatic PSFs generated by \textit{Kepler}'s optical model show Chromatic Aberration in the PSF and the sampling of the images by the CCDs pixel grid. Each of the sub-sampled images is shown on an area of 11x11 pixels; the intensity is inverted (dark is higher intensity) and histogram-equalization scaled. The central pixel can contain up to 50\% of the total intensity. Adapted from Fig.  14, \textit{Kepler} Instrument Handbook}
    \label{fig:kepler_psf}
\end{figure*}

As in many space-based observatories, especially wide-field ones, stellar images recorded by \textit{Kepler}'s CCD detector matrix are under-sampled; i.e. the images formed on the CCD surface contain spatial frequencies higher than the Nyquist sampling frequency, which is set by the spacing/size of the CCD pixels. In \textit{Kepler}, the ``plate scale'' is 3.98 arc seconds per pixel. While  stellar images (point spread function) have a 95\% enclosed energy diameter of 6.4 pixels, the central pixel can contain up to 50\% of the energy (i.e. Brightest Pixel Flux Fraction $\leq$ 0.5), which indicates a full width at half maximum of less than 2 pixels, the Shannon criterion for full sampling. The point spread function of the telescope can have spikes and other components of high spatial frequency (Fig.  \ref{fig:kepler_psf}), which exacerbate problems that arise from the under-sampling of stellar images.

\textit{Kepler} photometry has until now been based on aperture photometry using a so-called “pixel response function”, which is really a super-sampled point spread function of the entire system\cite{Bryson2010}, including optical effects of the telescope, pointing jitter and the underlying pixel response structure. This is now being improved in anticipation of photometry by fitting stellar profiles (``PRF fitting'') rather than photometry by simple integration of signal over an aperture\cite{Morris2014}. 

The outstanding precision of \textit{Kepler} aperture photometry, of about 10 parts in a million over 6 hours for 10th magnitude stars in its full original “K1” mission\cite{Christiansen2012} led to the anticipation that astrometry of 1 milliarcsecond accuracy might be possible, but in practice only about 4 milliarcsecond precision for a single measurement has been attained\cite{Monet2010}, with better precision being attained in certain cases. So far, centroid analysis of various kinds has been tried rather than PSF-fitting, but PRF fitting is now being implemented, although the astrometric benefits of this technique have not yet been published\cite{Morris2014}.

\subsection{K2 and Reduced Pointing Stability}

The unprecedented photometric precision achieved by \textit{Kepler} was largely made possible by the mitigation of many systematic errors having to do with the varying photometric sensitivity across the focal plane. These effects arise both from the optical design of the telescope (varying image quality across the large field of view) and from the pixel-to-pixel variations of the CCDs. \textit{Kepler} avoids both of these error sources by maintaining precise pointing as it orbits the sun.  The original \textit{Kepler} mission had a pointing stability (RMS jitter) of 0.063 pixels. The K2 mission has an increased RMS jitter of $\sim$0.5 pixels. This means that stars will drift across the pixels during and between exposures. Several advanced photometric techniques have been developed to deal with the increased systematic effects in \textit{K2} data, using codes such as K2SFF\cite{Vanderburg2014} and EVEREST\cite{Luger2016, Luger2018}. Whether the de-noising process is parametric or based on Gaussian Processes, detailed knowledge of the IPRF will be critical to mitigating the photometric errors associated with this drift. 

\section{\textit{Kepler} Data Analysis and the Pixel Response Function}

\textit{Kepler} image analysis has hitherto followed the general prescription set by Lauer (1999)\cite{Lauer1999}, which requires the determination of a super-sampled PSF, which is the convolution of the optical PSF of the telescope system and of the spatial sensitivity of a single pixel (Eq. \ref{eq:prf}). We call the spatial variability in sensitivity of a pixel the ``intrapixel response function'' (IPRF). 

\begin{figure*}
    \centering
    \includegraphics[width=\textwidth]{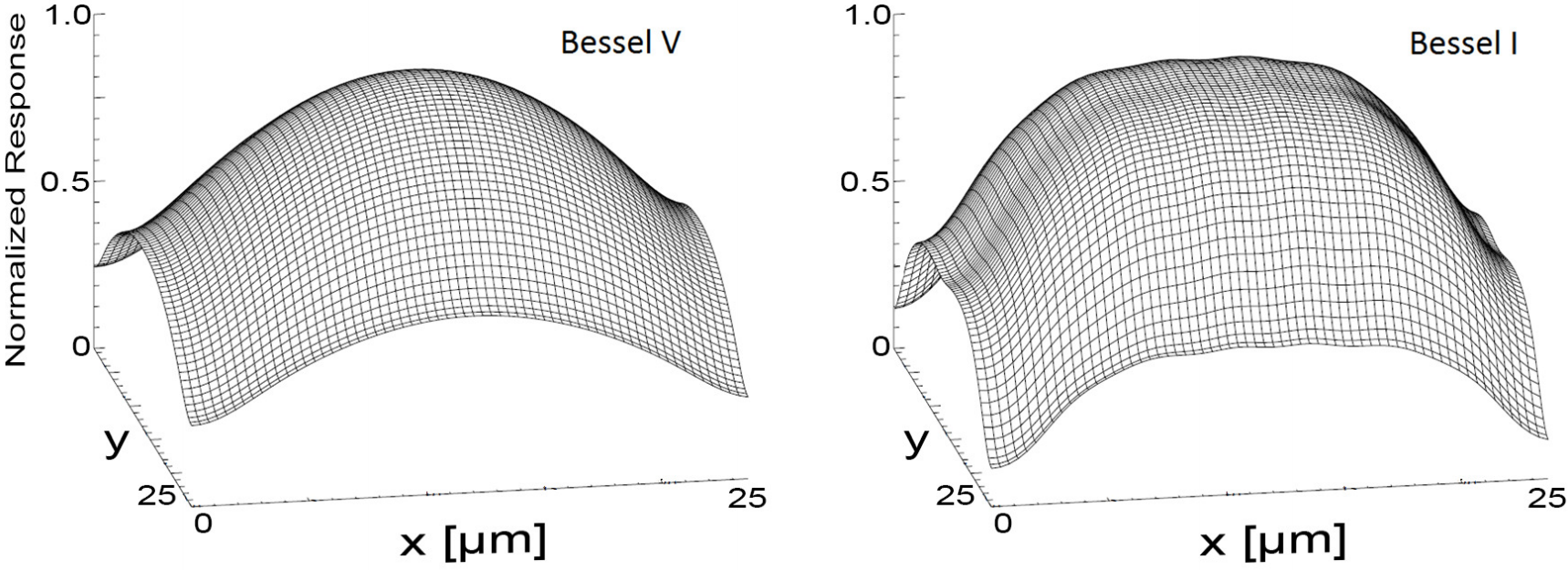}
    \caption{The IPRF of a SITe 502A backside illuminated CCD, with 25 $\times$ 25 $\mu m$ pixels measured in two broadband filters: Bessel V ($\sim$480 - 600 nm) and Bessel I ($\sim$720 - 1100 nm). In each case, the IPRF peaks near the physical center of the pixel and drops off rapidly towards the edges. However, the IPRF also depends significantly on the wavelength of light. Adapted from Piterman \& Ninkov 2002.}
    \label{fig:previous_iprf}
\end{figure*}

An ideal pixel has an IPRF that is spatially and temporally constant. The temporal stability of the intra-pixel variation has not been rigorously measured\footnote{Nevertheless, as part of this effort, we compared measurements of the IPRF made 4 months apart and find that they agree within the measurement uncertainties; see Section \ref{sec:iprfUniformity}.}; however, other characteristics of CCDs similar to those used for \textit{Kepler} have been shown to be extremely stable with time\cite{Koch2000}.  On the other hand, the spatial variability of the pixel response of 2\%-50\% has been measured for back-thinned CCDs\cite{Jorden1994,Piterman2002} (Fig.  \ref{fig:previous_iprf}). If left uncalibrated, this uncertainty in pixel response creates large errors in photometry and astrometry\cite{Bryson2010}. 

\subsection{The Intra-Pixel Response Function}

If one assumes that the intra-pixel sensitivity over a CCD is the same for each pixel (Jorden et al. 1994) and that it can be precisely determined, the intra-pixel response function can be separated from the optical PSF of the telescope.  (To avoid confusion, we will refer to the intra-pixel response function as the IPRF). Note that the IPRF can actually extend over several pixels, because the signal generated within one pixel can spill over to neighboring pixels; this can be caused by electron leakage and by the leakage of photons caused by a rapidly converging beam. 

Following Lauer (1999)\cite{Lauer1999}, we can write the super-sampled PSF or ``PRF'' as:
\begin{equation}
\label{eq:prf}
    PRF(x,y) = PSF(x,y)*IPRF(x,y)		 
\end{equation}

where $PSF$ is the ``optical'' PSF (including jitter),  $IPRF$ is the intra-pixel response function and $*$ the convolution operator. Spillage effects between pixels should be included in IPRF, which in fact spans a number of pixels, if necessary.

\section{Direct Measurement of the IPRF}
If the IPRF is well-characterized, the PSF can be precisely determined for each star in a field using knowledge of the optics and further refined with actual observations. A time-consuming raster of on-orbit observations to measure the ``PRF'' would no longer be required and the under-sampling effects could be corrected systematically, assuming that $IPRF(x, y)$ is the same for all pixels. 

The IPRF has been measured for frontside-illuminated\cite{Jorden1994, Kavaldjiev1998} and backside-illuminated devices\cite{Jorden1994, Piterman2002}. Typically, a single spot with a diameter which is smaller than the size of a pixel ($\sim$ 1 - 5 $\mu m$) is rastered across a set of several pixels to produce an X-Y grid with spacing much smaller than the size of a pixel. The measurements are then interpolated to estimate the IPRF (Fig.  \ref{fig:previous_iprf}).

\subsection{Measurement Setup}
\begin{figure*}
    \centering
    \includegraphics[width=0.8\textwidth]{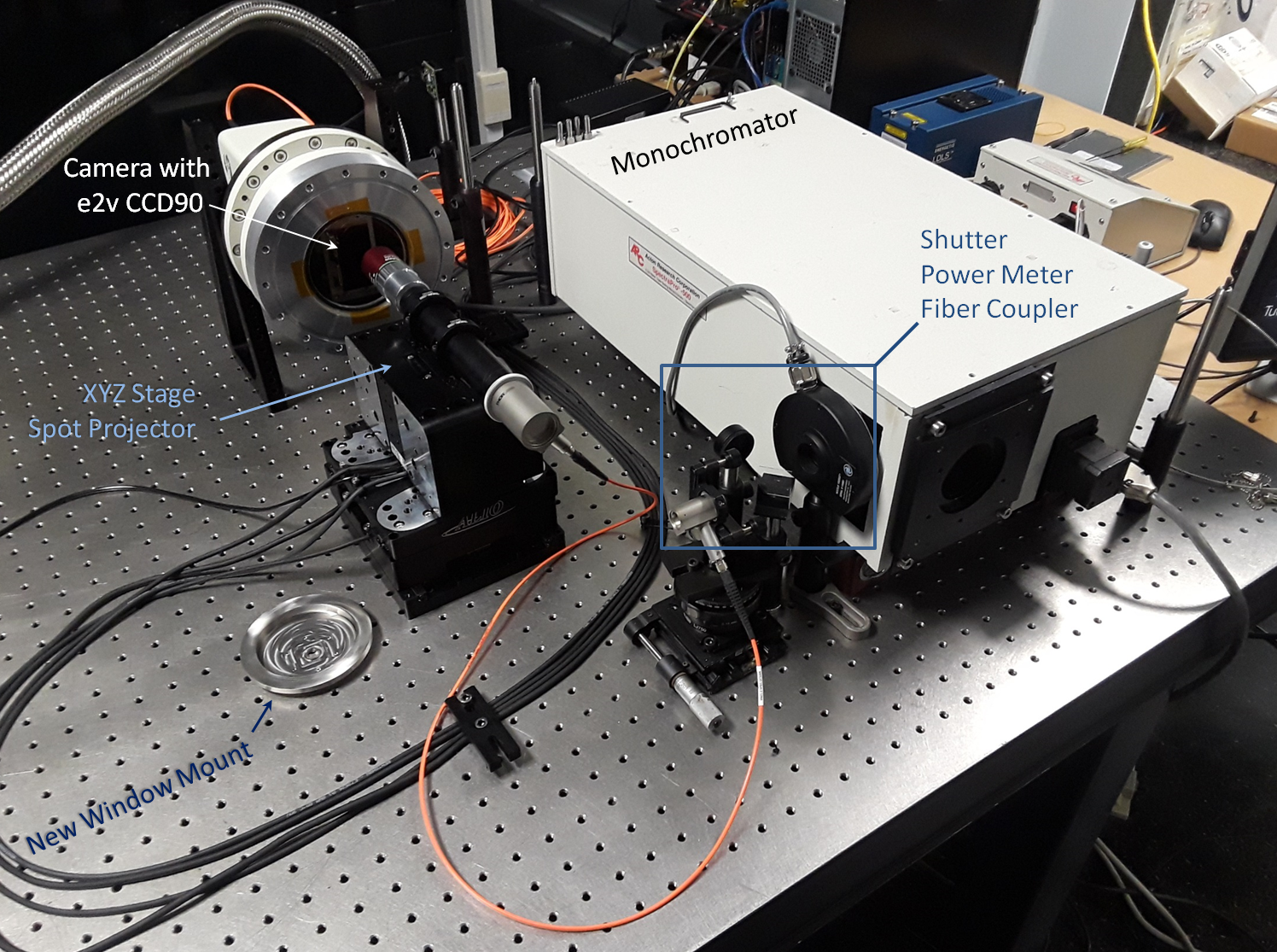}
    \caption{A new apparatus to measure the intra-pixel response function of a range of imaging arrays was built in RIT's Laboratory for Advanced Instrumentation Research. This system is able to produce small spots of light, across a wide range of wavelengths.}
    \label{fig:spotscanner}
\end{figure*}

A technique for the direct measurement of intra-pixel sensitivity variations has been developed and used at Rochester Institute of Technology for a number of years \cite{Kavaldjiev1998, Piterman2002}. Building on our previous experience, we built a new high performance measurement apparatus dedicated to the measurement of the IPRF (Fig.  \ref{fig:spotscanner}). This system is capable of producing spots with full width at half maximum (FWHM) of 3 microns or less (see Section \ref{sec:spotCharacterization}), at spacing intervals as small as 0.1 $\pm$ 0.02 $\mu m$. This allows us to sample \textit{Kepler}'s 27 $\mu m$ pixels with a wide range of sampling grids - from coarse (10 $\times$ 10) to extremely fine (270 $\times$ 270). A brief description of the major components is given below. 

\subsubsection{Light Source}
In this setup, we use an Energetiq EQ-99XFC Laser Driven Light Source, which produces a stable broadband spectrum (over the range 190 nm - 1000 nm, as implemented in our experiment). The source stability during a measurement is further monitored by a Thorlabs PM-100D power meter. Over the course of a typical measurement, the source output varies by $\sim$1.5\% rms on the scale of seconds; however, over the course of hours the instrument is stable to better than 0.1\%.

\subsubsection{Wavelength Selection}
\begin{figure*}
    \centering
    \includegraphics[width=0.8\textwidth]{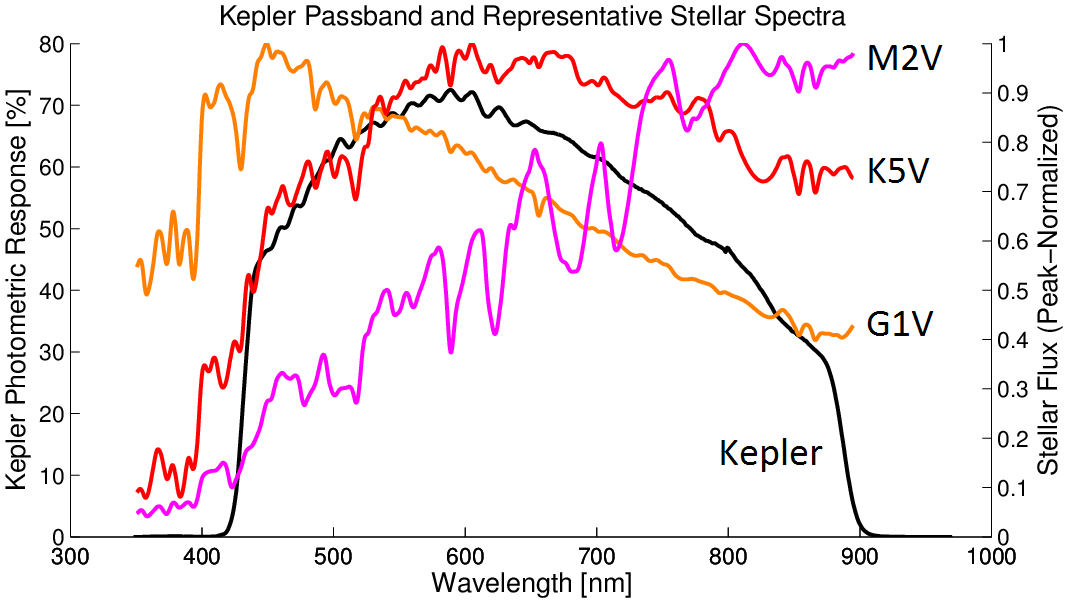}
    \caption{\textit{Kepler} observes stars of different colors through one passband. Both the optical PSF of the telescope and the IPRF change with wavelength. Therefore, calibrating the spectral dependence of the IPRF is extremely beneficial. Our measurement apparatus operates across the entire \textit{Kepler} passband.}
    \label{fig:stellarSpectra}
\end{figure*}

The broadband light source is filtered using an Acton Research Corportation 0.5 meter monochromator, to produce light with a spectral bandwidth of $\Delta \lambda\approx$15 nm. This allows us to characterize the dependence of the IPRF on wavelength. The wavelength dependence is especially important for the \textit{Kepler} photometer, due to its broad passband and the large variation in the spectral distribution of stars in the visible range (Fig.  \ref{fig:stellarSpectra}). Measurements in the spectral range longward of 700 nm are made by inserting a Bessel I filter at the entrance slit of the monochromator, to block the higher order UV and optical light.  

\subsubsection{Spot Projection} 
A wavelength-dependent characterization of \textit{Kepler}'s IPRF requires the ability to produce small-diameter spots across a wide range of wavelengths. For this purpose, we use a 20X Mitutoyo Plan Apo NIR infinity corrected objective. These long working distance objectives (20 mm) are well suited for illumination of sensors that must be housed in vacuum camera heads (such as the e2v CCD90). However, these fast objectives must be used with sufficiently thin windows, so as to not induce significant spherical aberration (see Section \ref{sec:windowEffects}). 

\subsubsection{Spot Translation}
\label{sec:spotTranslation}
The spot projector is fiber-fed and mounted on XYZ translation system from ALIO Industries. To be able to use a variety of cameras/dewars with the \textit{Kepler} CCD, we designed the system to translate the spot projector (rather than the sensor). The optical system is rigid to prevent sagging. The XYZ translation mechanism is capable of step sizes as small as 2 nm, with a bidirectional repeatability of 20 nm. This is an order of magnitude better than our minimum step size (250 nm). We chose this system for its excellent repeatability and thermal stability (the scale for the encoder is made from ZERODUR glass, yielding an expansion of 3 nm/K over a 30 mm distance). To reduce vibrational jitter, this apparatus is setup in a basement laboratory whose foundation is independent of the rest of the building.

\begin{figure}
    \centering
    \includegraphics[width=\linewidth]{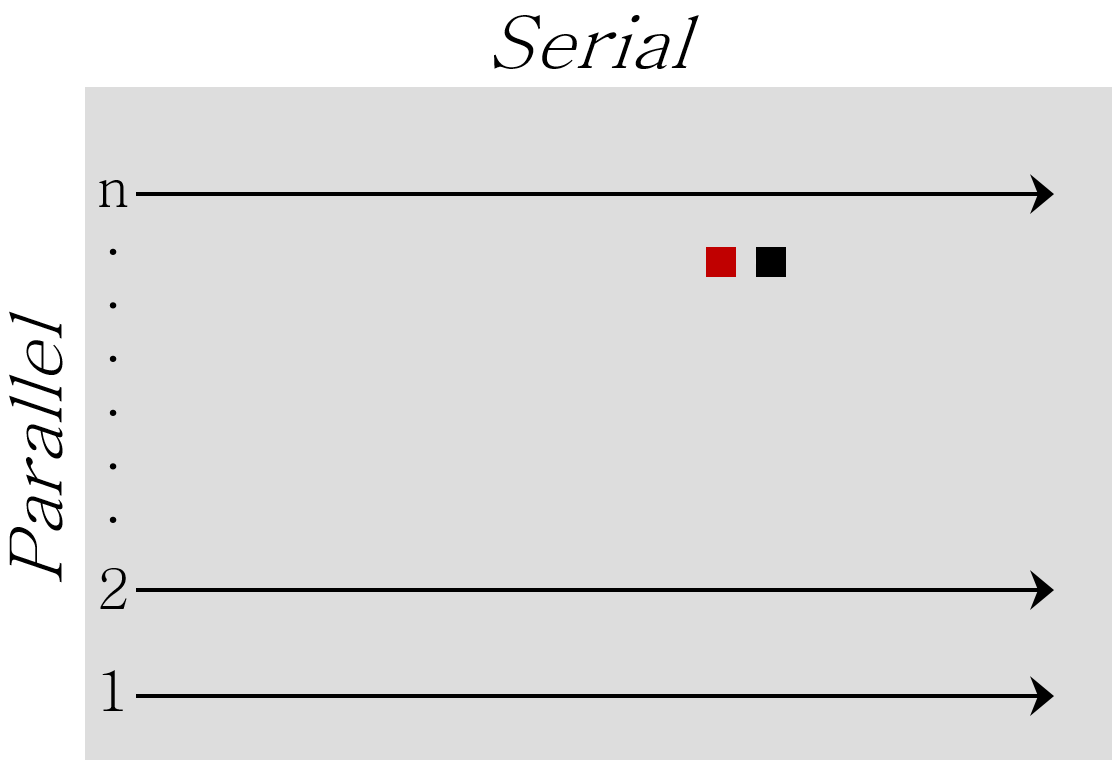}
    \caption{The 2D sampling grid was created by acquiring a set of measurements in the serial direction, advancing to the next parallel position, and acquiring another set of measurements in the serial direction, until the entire region of interested has been measured. The approximate locations of the pixels measured in the May 2018 and September 2018 campaigns are indicated by the red and black squares (see Sec. \ref{sec:iprfUniformity})}.
    \label{fig:scanSetup}
\end{figure}

We measured the IPRF of \textit{Kepler}'s CCD by scanning a 54 $\mu m$ $\times$ 54 $\mu m$ region of interest, with steps of 1 $\mu m$ in each direction. The measurements were performed at 50 nm intervals, over the range of 400 to 850 nm, (with a passband of $\Delta \lambda \sim 15$ nm). Before each measurement, the spot projector was refocused by scanning through a range of focus positions\footnote{In our system, the position of best focus changed in a monotonic fashion by a total distance of 20 $\mu m$ from 400 nm to 850 nm.}. The 2D sampling grid is achieved by successive linear scans in the serial direction (Fig.  \ref{fig:scanSetup}). For example, for a 10 $\times$ 10 sampling grid, 10 steps/measurements are are made in the serial direction, the spot projector is moved 1 step in the parallel direction, and another 10 steps/measurements are acquired in the serial direction, until the entire sampling region has been covered. Each scan was performed three times, to verify stability and improve the signal-to-noise ratio; a full scan was completed before the next began. Our initial scans, performed over a larger area (70 $\mu m$ $\times$ 70 $\mu m$), but at only a few wavelengths (450 nm, 600 nm, and 800 nm) are presented in a previous work\cite{Vorobiev2018}. 

\subsubsection{CCD Camera}
\label{sec:setupCCDCamera} 
The e2v CCD90 (SN 208) was acquired through NASA Ames Research Center, from Ball Aerospace Corporation. The sensor was installed in a Spectral Instruments 800 Series camera. This camera allows us to cool the sensor to -45$^{\circ}$ C, which is significantly warmer than \textit{Kepler}'s science operating temperature of -85$^{\circ}$ C. The temperature we reach in the lab (-45$^{\circ}$ C) reduces the dark current to a manageable level; however, the absorption length in silicon depends on temperature, as well as wavelength. Therefore, a wavelength-temperature correction may be needed to relate our measurements to \textit{Kepler}'s on-orbit performance (see Appendix \ref{sec:tempratureWavelengthEffects} for a further discussion).

\begin{figure*}
    \centering
    \includegraphics[width=\textwidth]{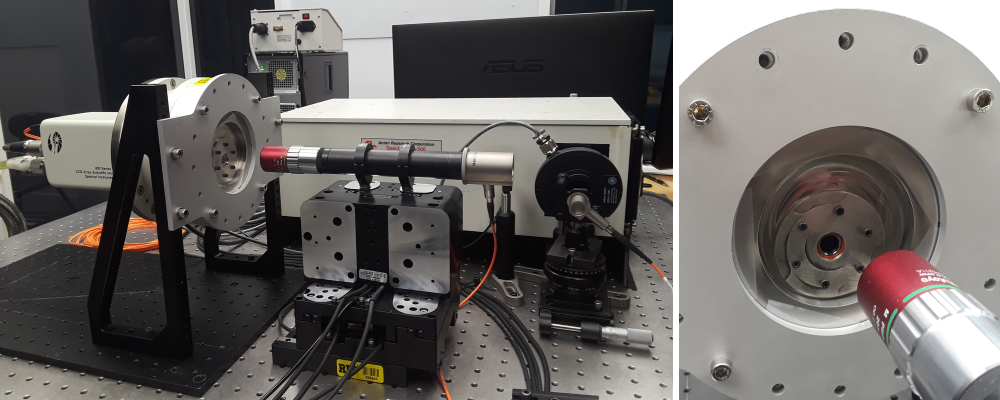}
    \caption{The camera's normal 4 inch window was replaced with a much smaller, thinner fused silica window, which is mounted in a stainless steel frame.}
    \label{fig:windowZoom}
\end{figure*}

The CCD camera's original 4 inch diameter window was replaced with a steel frame and a much smaller fused silica window. The new window is 500 $\mu m$ thick, with a diameter of 10 mm (Fig.  \ref{fig:windowZoom}). This significantly reduces the areas on the sensor which are accessible for measurement. However, the remaining area is sufficient to measure thousands of pixels. Furthermore, the window was placed off-axis, so that different regions of the sensor can be probed by rotating the mount. 

\begin{figure*}   
    \centering
    \includegraphics[width=\textwidth]{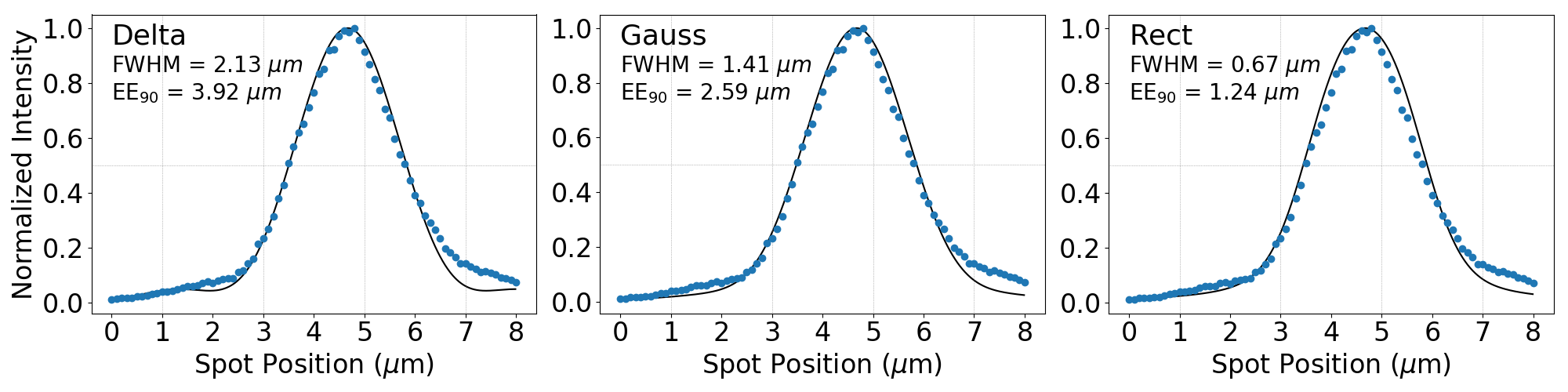}
    \caption{The spot size produced by the spot projector is estimated by fitting the measured profile (dots) with an Airy function convolved with one of three models for the pixel response function of the Canon 120MXS sensor: delta-like (Left); Gaussian (Middle); and uniform (Rect, Right). The size of the reconstructed Airy function is characterized by the full-width at half-maximum (FWHM) and the size of a region that encloses 90\% of the total energy of the spot (EE$_{90}$). Note that assuming a perfectly uniform response (Rect) for the Canon pixels can result in impossibly small spot sizes. These results are for 700 nm light.}
    \label{fig:spotSize_noWindow}
\end{figure*}
\subsection{Spot Characterization}
\label{sec:spotCharacterization}


The spot size was measured using a Canon 120MXS CMOS sensor: a front-illimunated imager with 2.2 $\mu m$ pixels and microlenses. This device is well suited for spot-profiling, due to its small pixel size and low read noise ($\sim$2.3 e$^-$ rms), which helps maintain a high SNR in the ``wings'' of the point-spread function. The spot size was measured at 50 nm intervals, over the range from 400 nm to 850 nm. The Canon sensor's 2.2 $\mu m$ pixels are still too large to sufficiently sample a spot with a FWHM of $\sim$3 $\mu m$ using conventional imaging. Therefore, we scanned the spot across a reference pixel, using 0.1 $\mu m$ steps, to create a super-sampled intensity profile (Fig.  \ref{fig:spotSize_noWindow}). These profiles are a convolution of the optical spot shape, the pixel response function of the Canon sensor, and the step size / sampling rate. Our simulated observations showed that a 0.1 $\mu m$ step was sufficiently small to accurately reproduce the intensity profile of the optical spot. 

To estimate the actual size of the optical spot, we modeled the measured response as a convolution of an Airy function and one of several pixel response functions, which include Gaussian diffusion of carriers within the pixel (Fig.  \ref{fig:spotSize_noWindow}). The three models to describe the pixel response function were:
\begin{enumerate}
    \item A delta-like pixel response (only sensitive in the center)
    \item A peaked response in the center with a Gaussian fall-off of $\sim50\%$ towards the edges
    \item A perfectly uniform response across the entire pixel area (Rect).
\end{enumerate}

Because the 120MXS sensor is front-illuminated and uses microlenses, we expect the Gaussian PRF to be the best approximation, with the two edge cases being the upper and lower limits on spot size, respectively, for the Delta and Rect PRFs. In our analysis, we used the spot size (at each wavelength) as estimated by assuming a Gaussian PRF for the Canon 120MXS sensor.

\subsubsection{Effects of the Camera Window on Spot Size}
\label{sec:windowEffects}
\begin{figure}
    \includegraphics[width=\linewidth]{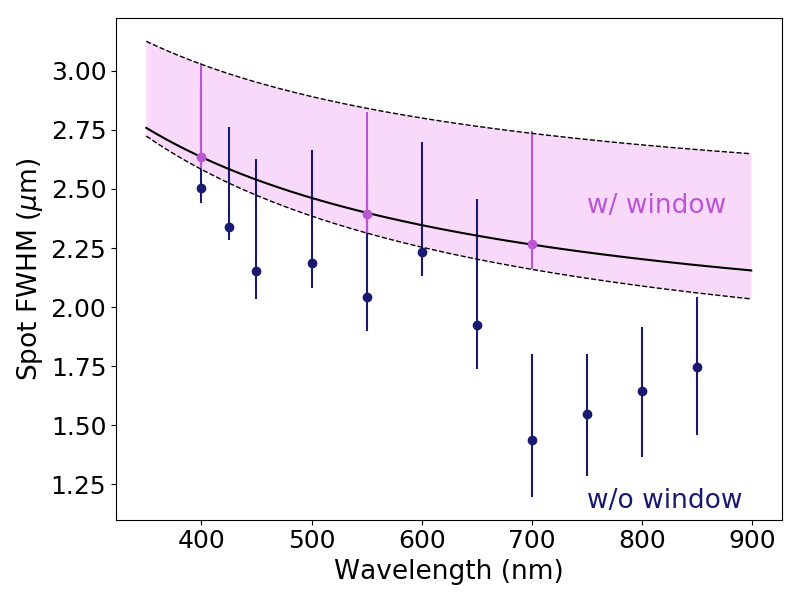}
    \caption{The size of the optical spot was measured at several wavelengths, using a Canon 120MXS CMOS sensor with 2.2 $\mu m$ pixels, using the procedure described in Section \ref{sec:spotCharacterization}. The error bars represent the upper and lower limits, given by assuming a delta-like and uniform PRFs for the Canon pixels, respectively. These measurements were repeated after introducing an extra 500 $\mu m$ thick window into the beam, at 400, 550, and 700 nm.}
    \label{fig:spotProfile_withWindow}
\end{figure}

The e2v CCD90 at the focus of this work is housed in a vacuum camera head, with a 500 $\mu m$ thick fused silica window. Some degradation of the optical performance of the Mitutoyo objective is expected, due to the spherical aberration which arises when a window is introduced into the quickly converging beam. To estimate the magnitude of this effect, the measurement described in the previous section (Sec. \ref{sec:spotCharacterization}) was repeated, with an identical 500 $\mu m$ window in the beam. The system was refocused (to account for the change in focal distance). The resulting spot measurement is shown in Fig.  \ref{fig:spotProfile_withWindow}. 

The spot measured with the window is 2 - 3 $\mu m$ FWHM, depending on the wavelength and choice of PRF for the Canon sensor. The introduction of a window into the beam appears to degrade the optical performance of the spot projector and increases the size of the spot. The degradation is worse in the infrared, where the objective's performance is otherwise optimal. Because we did not re-characterize the system's performance at all wavelengths of interest, we estimate the ``degraded'' spot size by fitting smooth functions to the spot size measurements at 400, 550 and 700 nm, and to the lower and upper bounds, separately. 

To complicate matters further, the Canon sensor already has a thin window near the detector's surface and the Mitutoyo objective is designed to be used with thin microscope cover slides. However, it is not clear to us (and we did not investigate) if the placement of the window (in the middle of the beam vs. near the focal plane) has an effect on the spot size. Instead, we take a conservative approach and use the larger ``with window'' spot size estimations in our analysis. As we describe in the following section, even these larger spots have a minor (but not negligible) effect on the estimation of \textit{Kepler}'s IPRF. 

\subsubsection{Effects of Spot Size on the IPRF Measurement}
\label{sec:spotSizeEffects}
\begin{figure*}
    \centering
    \includegraphics[width=\textwidth]{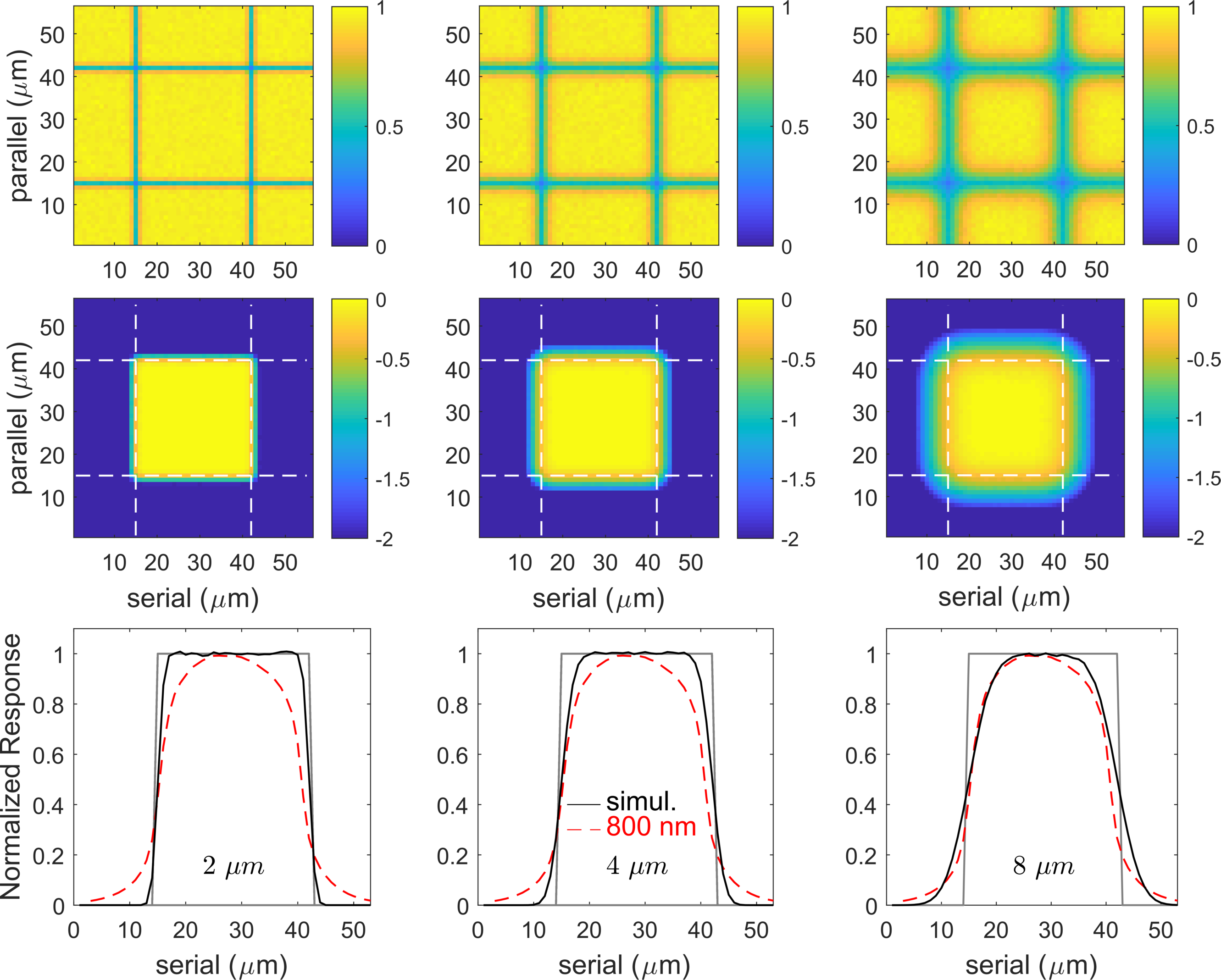}
    \caption{In these spot scan experiments, the measured response can be dominated by blurring effects due to a too large spot. We simulated our spot scan experiment using a pixel with uniform IPRF, and Gaussian spots with FHWM of 2, 4, and 8 $\mu m$. Top: raw output from a simulated scan, peak-normalized; the ``probability'' of a photon being recorded in a particular pixel, as a function of position. Middle: The response (log scale) of a single pixel, as a function of spot position. Bottom: Horizontal profiles (mean-normalized) across the maps shown in the Middle row, comparing the results of a simulated scan of a ``perfect pixel'' (gray line) and our actual measurement at 800 nm; in our actual experiment, the spot sizes vary from $\sim$2.75 $\mu m$ - 2 $\mu m$, depending on the wavelength (see Fig.  \ref{fig:spotProfile_withWindow}).}
    \label{fig:effects_of_spot_size}
\end{figure*}

The spot scanning technique works best when the spot illuminating the pixel is much smaller than the size of the pixel - ideally, the spot would be infinitesimal. As the spot size increases, measurements near the edge of the pixel will be compromised by light spilling into neighboring pixels, mimicking the effects of diffusion that one hopes to characterize as part of the IPRF. To estimate the magnitude of this effect, we performed a series of simulated spot scans, using a pixel with a perfect IPRF (uniform response across the pixel, without any diffusion of carriers) and the level of Poisson noise observed in actual scans. The pixel size was 27 $\mu m$, to match the \textit{Kepler} CCD, and the spots were modeled as Gaussian distributions with FWHM of 2, 4 and 8 $\mu m$. The resulting raw IPRF measurements are shown in Fig.  \ref{fig:effects_of_spot_size}.

Our simulated spot scans show that the blurring of the measured IPRF near the pixel's edge due purely to the geometry of the illuminating spot is easily resolved, even for spots with a FHWM of 2 $\mu m$. This effect is manageable in our measurements, because the CCD90 pixels are quite large. In general, we estimate that spots with FWHM $<\frac{1}{4}$ of the pixel size can be used, without introducing too much ambiguity into the measured response (see Appendix \ref{sec:deconvolvingIPRF}). For example, the effects of diffusion seen in the raw measurement at 800 nm are still easily distinguished from the blurring induced by a 4 $\mu m$ spot, but not by an 8 $\mu m$ spot; the IPRF shows less diffusion at the longer wavelengths, making this the more challenging comparison. In addition to using spot sizes $<$ $4\mu m$ in our measurements of the IPRF, we attempt to account for this blurring using Lucy-Richardson de-convolution (for details, see Appendix \ref{sec:deconvolvingIPRF}).

\subsection{Measurement Results}
We measured the IPRF of \textit{Kepler}'s CCD by scanning a 54 $\mu m$ $\times$ 54 $\mu m$ region of interest, with steps of 1 $\mu m$ in each direction, using the procedure detailed in Section \ref{sec:spotTranslation}. The measurements were performed at 50 nm intervals, over the range of 400 to 850 nm, (with a passband of $\Delta \lambda \sim 15$ nm). Before each measurement, the spot projector was refocused by scanning through a range of focus positions\footnote{In our system, the position of best focus changed in a monotonic fashion by 20 $\mu m$ from 400 nm to 850 nm.}. Each scan was performed three times, to verify stability and improve the signal-to-noise ratio; a full scan was completed before the next began.  The 2D sampling grid is achieved by successive linear scans in the serial direction. For example, for a 10 $\times$ 10 sampling grid, 10 steps/measurements are are made in the serial direction, the spot projector is moved 1 step in the parallel direction, and another 10 steps/measurements are acquired in the serial direction, until the entire sampling region has been covered. Our initial scans, performed over a larger area (70 $\mu m$ $\times$ 70 $\mu m$), but at only a few wavelengths (450 nm, 600 nm, and 800 nm) are presented in a previous work\cite{Vorobiev2018}. 

\begin{figure*}
    \centering
    \includegraphics[width=\textwidth]{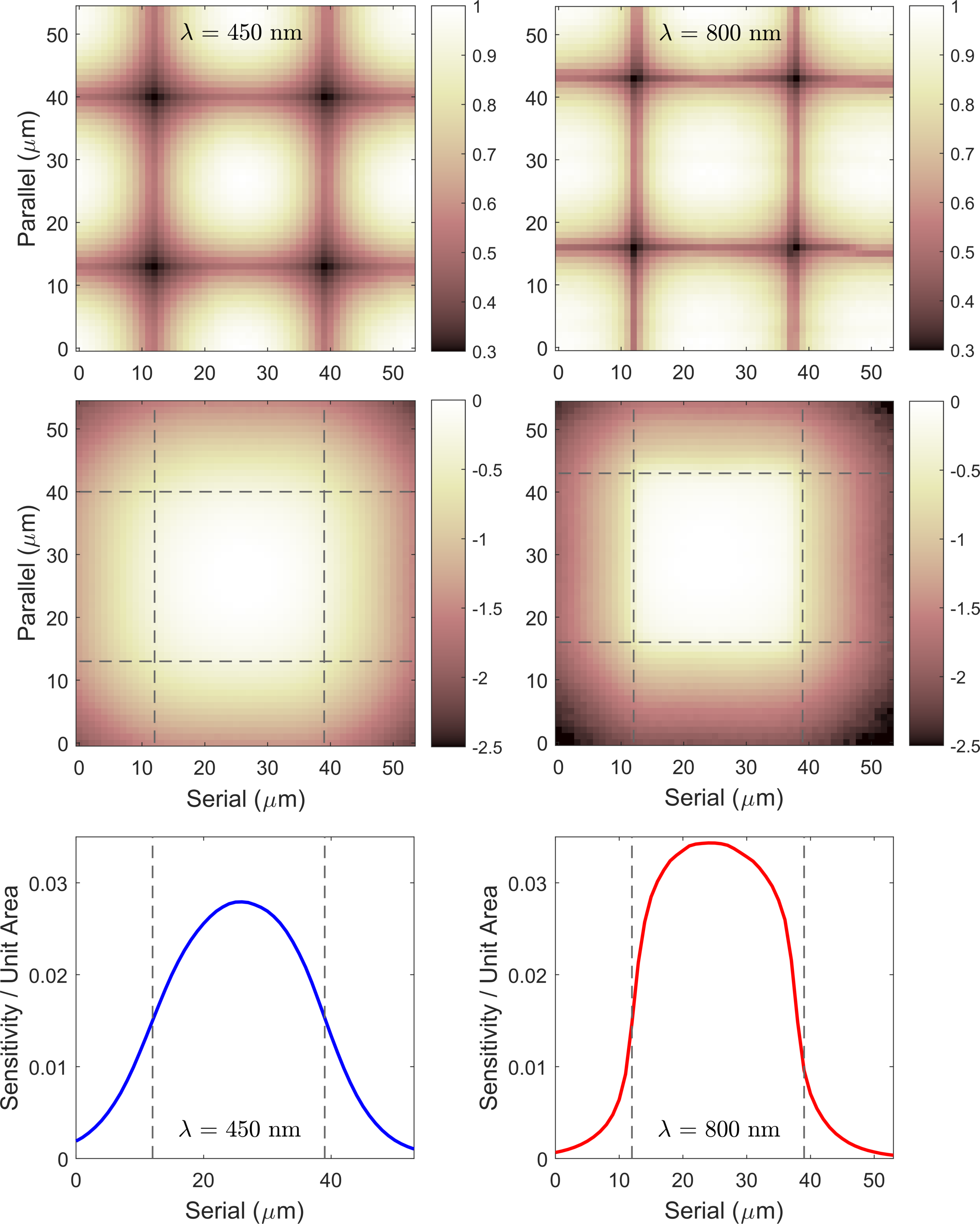}
    \caption{The raw results of the spot scanning experiment for \textit{Kepler}'s CCD at 450 nm and 800 nm. Top: the raw response of the brightest pixel, as a function of spot position. Middle: the normalized response of a single reference pixel, as a function of spot position. Bottom: a normalized profile of the single pixel response shown in the middle row. In general, the response is maximal the central third of a pixel, and rapidly decreases towards the pixel edges.}
    \label{fig:450nm_fullScan}
\end{figure*}

The raw results of an entire scan at 450 nm and 800 nm are shown in Fig.  \ref{fig:450nm_fullScan}. These maps were created by recording the maximum pixel value, as a function of spot position, and normalizing by the brightest value observed. This is analogous to the probability that a photon incident on a certain location within a pixel has to be measured by that pixel. The structure in Fig.  \ref{fig:450nm_fullScan} is largely due to the diffusion pattern of carriers in Kepler's CCD, with a small contribution from the finite size of the spot (Section \ref{sec:spotCharacterization}) used to illuminate the pixel.

\subsection{\textit{Kepler}'s IPRF vs. Wavelength}
\begin{figure*}
    \centering
    \includegraphics[width=\textwidth]{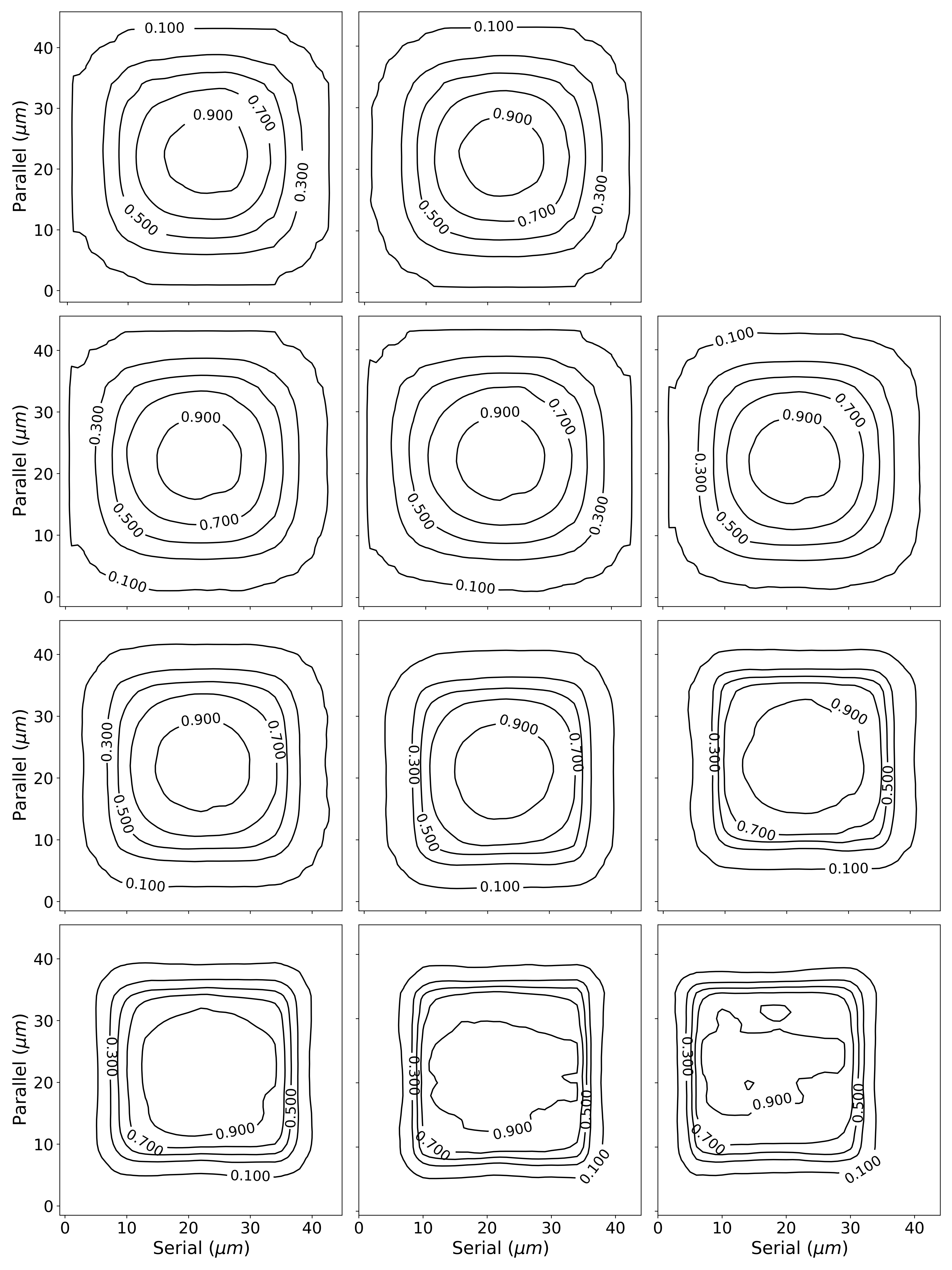}
    \caption{The measured intra-pixel response function (IPRF) for the \textit{Kepler} CCD90 sensor, over the wavelength range of 400 to 850 nm. Each map is peak-normalized, to emphasize the dependence of the IPRF shape on wavelength.}
    \label{fig:kepler_iprf_vs_wavelength}
\end{figure*}

We measured the IPRF of the CCD90 sensor for multiple wavelengths across the \textit{Kepler} passband. The 2D response maps obtained for each wavelength are shown in Fig.  \ref{fig:kepler_iprf_vs_wavelength}). These maps are made by peak-normalizing the response of a single pixel, as a function of spot position (similar to the middle row of Fig.  \ref{fig:450nm_fullScan}). To account for the blurring related to the spot size, we performed Lucy-Richardson de-convolution on the raw measurements (see Appendix \ref{sec:deconvolvingIPRF} for details). For each wavelength, we used the appropriate size spot (Fig.  \ref{fig:spotProfile_withWindow}), estimated using the Gaussian diffusion model for the Canon sensor (Sec. \ref{sec:spotCharacterization}).

\begin{figure*}
    \centering
    \includegraphics[width=\textwidth]{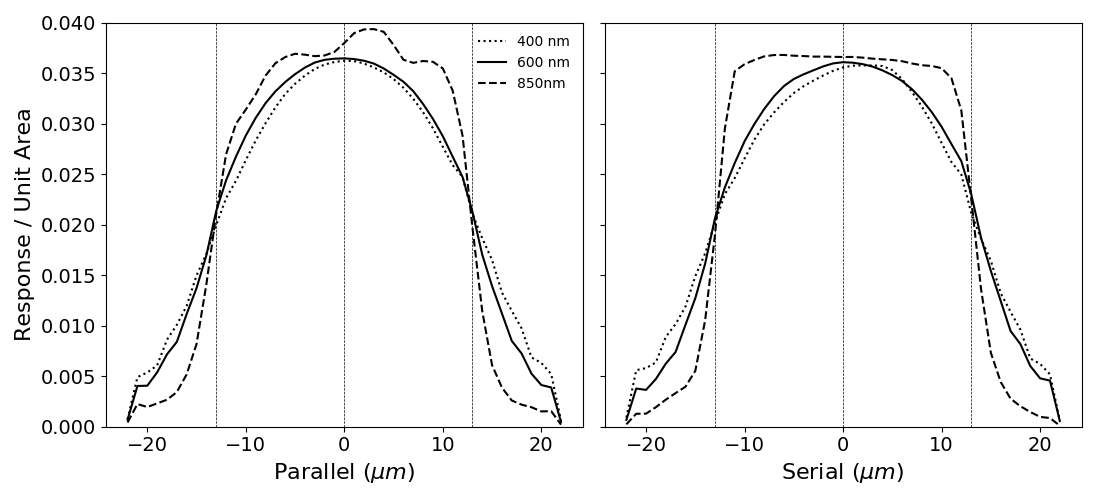}
    \caption{The measured intra-pixel response function (IPRF) for the \textit{Kepler} CCD90 sensor, for several representative wavelengths across the \textit{Kepler} passband,  shown as profiles based on 2D maps that are volume-normalized. Note that the response in the parallel direction (Left) is different from the response in the serial direction (Right), especially for longer wavelength light.}
    \label{fig:parallelSerialProfiles}
\end{figure*}

The IPRF shape shows a strong dependence on wavelength, across \textit{Kepler}'s passband. At the shortest wavelengths, the IPRF exhibits smooth azimuthal response near the center, with a gradual fall-off towards the pixel edges. As the wavelengths get longer, the IPRF becomes more defined and square; at the longest wavelengths, the electric field created by the gate structure on the reverse side of the CCD starts to become apparent. This behavior is consistent with that of other back-illuminated sensors. The shorter wavelength light is absorbed closer to the surface, further from the gates, and experiences Gaussian-like diffusion. Conversely, the photons that are absorbed closer to the gates are much better confined in the depletion region, which results in a more steep, better defined IPRF. The profiles in the parallel and serial directions for several representative wavelengths are shown in Fig.  \ref{fig:parallelSerialProfiles}. These volume-normalized profiles can be compared directly to each other, as opposed to peak-normalized plots (which can only be used to make relative comparisons regrading the IPRF shape). 

\subsection{Spatial Uniformity of the IPRF}
\label{sec:iprfUniformity}
At some level, the IPRF varies from pixel to pixel on small and large scales, due to the nonuniformity of the fabrication process. The electric field structure within a pixel (which largely defines the IPRF) depends primarily on the doping uniformity in the silicon layers, the thickness/quality of the insulator layer and the physical structure of the gate electrodes. In general, scientific grade CCDs offer excellent spatial uniformity. For example, the pixel-to-pixel response non-uniformity for these sensors was measured to vary at a level of 1$\sigma=1\%$ (see Section 4.12 of the Kepler Instrument Handbook\cite{vancleve2016}).

\begin{figure*}
    \centering
    \includegraphics[width=1\textwidth]{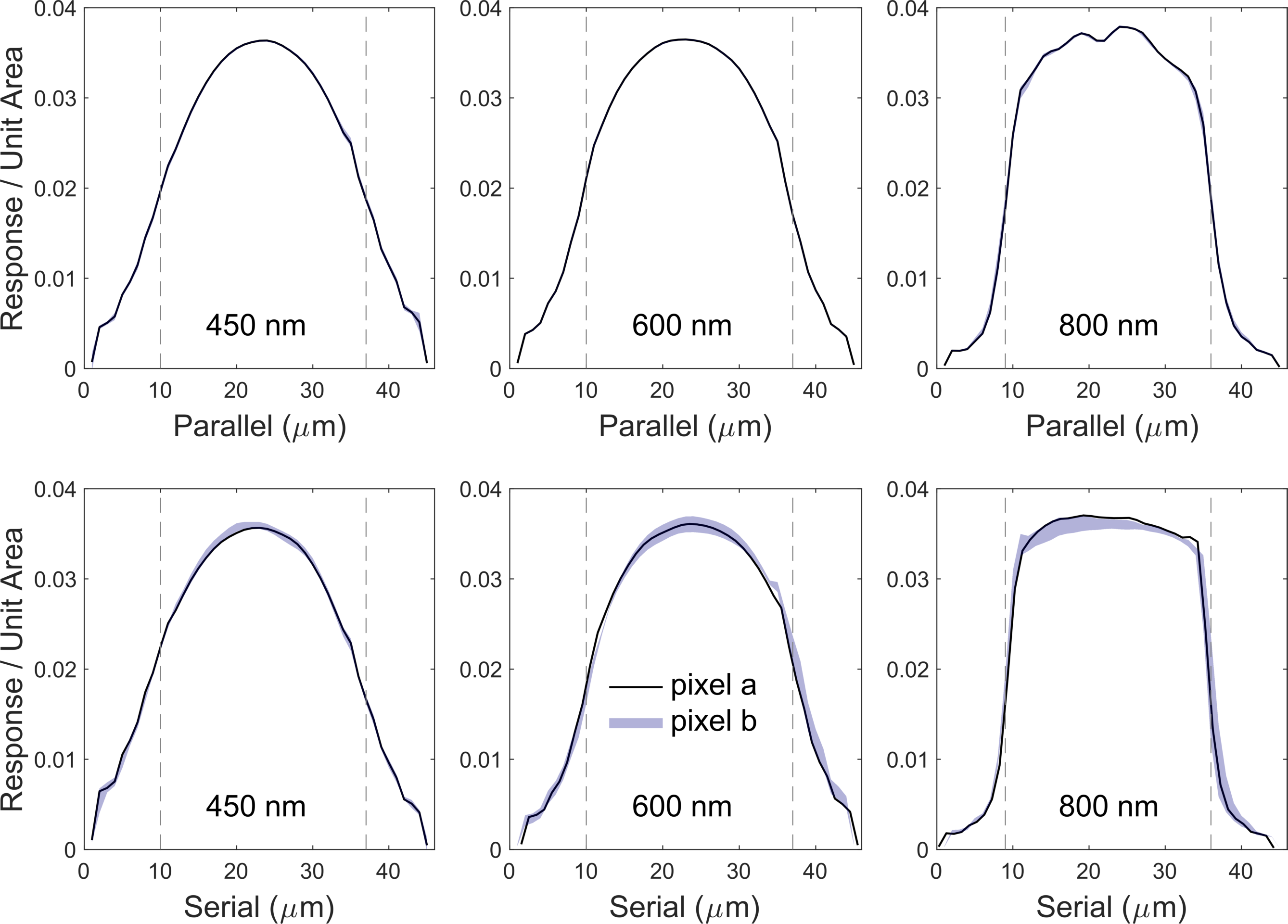}
    \caption{The IPRF does not appear to show significant variability from pixel-to-pixel. The IPRFs measured for ``pixel a'' (May 2018) and ``pixel b'' (September 2018), which are separated by 1.566 mm, appear to agree. Here, the shaded area indicates the upper and lower bounds (a 2$\sigma$ range) for the IPRF estimated from a set of 3 measurements in September. The agreement is best in the parallel direction, where the measurements exhibit less scatter overall. The measurements in the serial direction are acquired first, before advancing in the parallel direction and acquiring another set of measurements in the serial direction. The dashed vertical lines demarcate pixel boundaries.}
    \label{fig:iprf_nonuniformity}
\end{figure*}

Although our camera window can be rotated to sample multiple locations on the sensor, we did not perform a systematic study of the IPRF non-uniformity as part of this effort. To estimate the variation in the IPRF from pixel-to-pixel, we compare the results acquired in May, 2018\cite{Vorobiev2018} on ``pixel a'', to those acquired in September, 2018 on ``pixel b''. The two pixels are located at row, column positions of 371, 1386 and 370, 1444 respectively, resulting in a physical separation of $\sim1.566$ mm (see Fig.  \ref{fig:scanSetup}). The IPRFs obtained at each of these locations are shown in Fig.  \ref{fig:iprf_nonuniformity}. The shaded regions in the plot indicate the upper and lower bounds on the IPRF estimated using the sets of three measurements acquired at each wavelength during the September 2019 campaign (See Section \ref{sec:setupCCDCamera}. Specifically, the bounds were defined using the following relations: $upper$ $bound = median + standard$ $deviation$ and  $lower$ $bound = median - standard$ $deviation$. 

Generally, the IPRFs obtained for ``pixel a'' and ``pixel b'' are consistent, given the repeatability of each measurement. The agreement is best in the parallel direction, with more scatter in the serial direction. This preliminary result supports the hypothesis that the IPRF is not likely to vary significantly across the CCD (or from one sensor in the focal plane to another), given the high degree of uniformity for Kepler's devices. This is further supported by previous measurements of IPRF, which also found little variability from pixel to pixel\cite{Kavaldjiev1998}.

\subsection{The \textit{Kepler} Response and the Integrated IPRF}
\begin{figure}
    \centering
    \includegraphics[width=\linewidth]{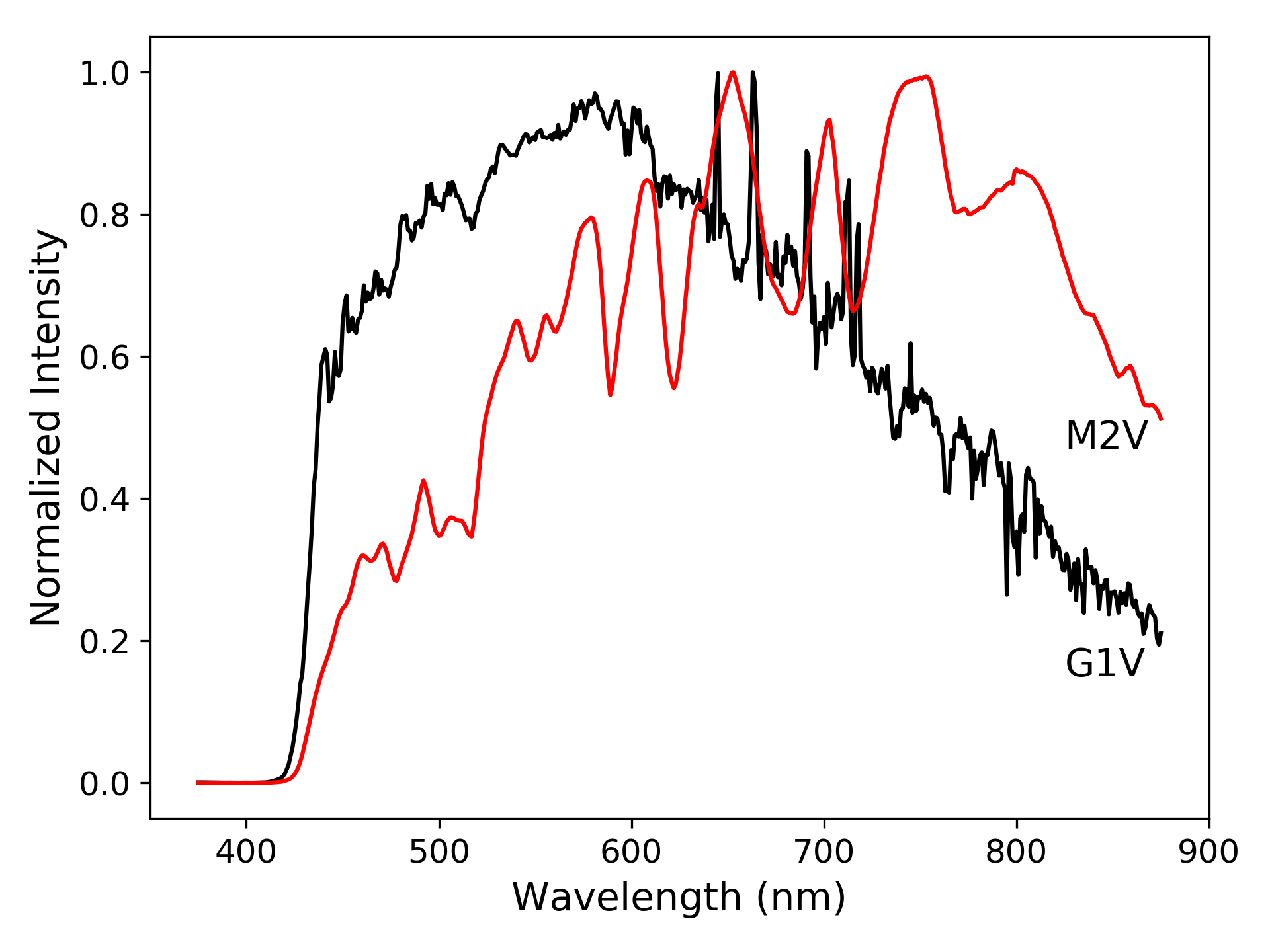}
    \caption{The spectral energy distribution of different spectral types convolved with \textit{Kepler}'s spectral response should be accounted for when constructing a combined ``effective'' IPRF from the spectral IPRF measurements. Note that the longer wavelength measurements show higher frequency structure that is not seen at shorter wavelengths; most likely, this traces the electric field structure within the pixel, which is more pronounced deeper in the device, closer to the gates.}
    \label{fig:MG_spectra}
\end{figure}

\textit{Kepler} photometry is performed over a single, large passband that covers the range of $\sim$425 to 850 nm (Fig.  \ref{fig:stellarSpectra}). In order to use the spectral IPRFs that we measured as part of this work (Fig.  \ref{fig:kepler_iprf_vs_wavelength}) in a data reduction pipeline, we identify three issues that must be addressed:

\begin{enumerate}
    \item The spectral IPRFs must be combined into a single, integrated ``effective'' IPRF.
    \item The spectral energy distribution of stars varies across the \textit{Kepler} passband (Fig.  \ref{fig:stellarSpectra} and Fig.  \ref{fig:MG_spectra}).
    \item \textit{Kepler's} point-spread function is chromatic (Fig.  \ref{fig:kepler_psf}) and the intensity distribution at the focal plane depends on a star's spectral type.
\end{enumerate}

\begin{figure*}
    \centering
    \includegraphics[width=\textwidth]{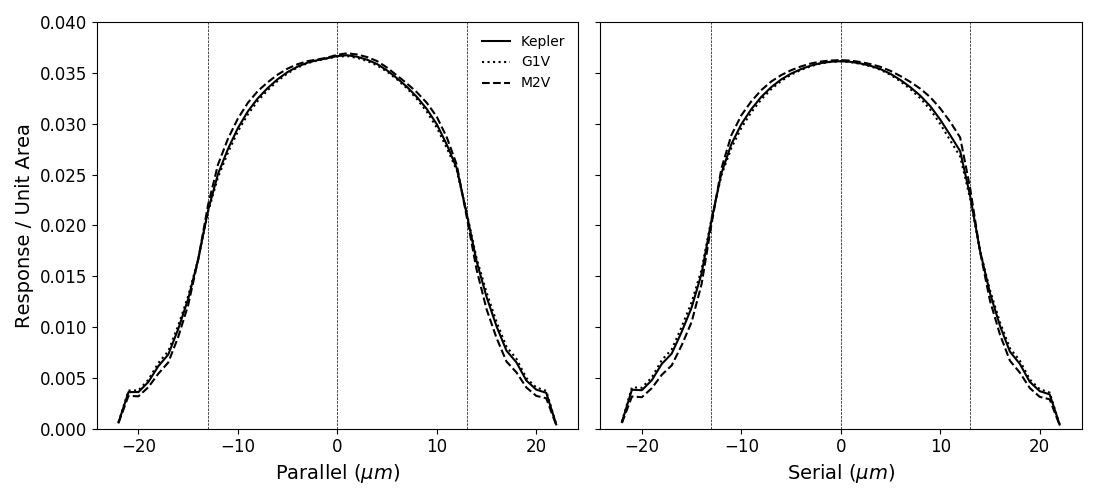}
    \caption{The integrated IPRF of the \textit{Kepler} photometer based on \textit{Kepler}'s spectral response only (solid line) and the IPRF calculated for stars with two representative spectral types, a G dwarf (dotted line) and an M dwarf (dashed line). These profiles are volume-normalized over the scan region.}
    \label{fig:combined_iprfs}
\end{figure*}

The tasks outlined above are listed in order of increasing complexity (and presumably improving fidelity). To obtain the integrated IPRF, we combine the spectral IPRFs using a mean, weighted by \textit{Kepler's} spectral response, at the wavelengths of interest. To go one step further, we also weighted the IPRF by the spectral energy distributions of a G dwarf and M dwarf, as seen by \textit{Kepler} (Fig.  \ref{fig:MG_spectra}). The resulting IPRFs are shown in Fig.  \ref{fig:combined_iprfs}). The profiles shown here are volume-normalized over the scan region, to allow a direct comparison between the different IPRFs; note that a small difference here may translate to a large photometric difference, once the response per unit area is integrated over the pixel area. 

\section{Discussion of Results}
We have presented the design and the results of our effort to directly characterize the intra-pixel response function of the e2v CCD90 sensor, across the passband of the \textit{Kepler} photometer. We have shown that our spot projector is able to generate spots of FWHM $\lesssim$3 $\mu m$, over the spectral range 400 - 850 nm. The source and the mechanical translation system appear quite stable, as can be inferred from the very consistent response measured over the course of several hours (Fig.  \ref{fig:450nm_fullScan}) and between successive scans.

The IPRF we measured for \textit{Kepler}'s CCD appears similar to the IPRF measured for other back-illuminated CCDs\cite{Jorden1994, Piterman2002} (Fig.  \ref{fig:previous_iprf}). Two main trends can be seen in these initial data: 

\begin{enumerate}
    \item The IPRF varies significantly between the center and edges of a pixel. The sensitivity at the edge of a pixel is $\sim$50\% lower than at the pixel's center, and as much as 70\% lower near the pixel corner.
    \item The IPRF is a function of wavelength, as well as position. The IPRF in the near-IR is noticeably different than in the visible range. 
\end{enumerate}

\subsection{Applying the Measured IPRF to \textit{Kepler} and \textit{K2} Data}
\label{sec:kepler_iprf}
The spectral IPRFs measured in our lab must be combined into a broadband ``effective'' IPRF, before the IPRF can be used in a data reduction pipeline. We recommend combining the individual IPRFs using a weighted average, where the weights are provided either by \textit{Kepler}'s spectral response or by the spectral energy distribution of the target star convolved with \textit{Kepler}'s response. Although the differences between these IPRFs may appear small when shown as volume-normalized profiles (Fig.  \ref{fig:combined_iprfs}), we expect the differences will be significant when the response is integrated over the pixel area. 

The raw measured IPRFs obtained with the spot scanner contain some contribution from the finite size of the spot. To estimate the ``un-blurred'' IPRF, which is due only to carrier diffusion within the pixel, we perform Lucy-Richardson de-convolution on the raw 2D pixel response maps, using our estimation of the spot size at each wavelength (Fig.  \ref{fig:spotProfile_withWindow}). We expect this to be the more true approximation of the actual IPRF; however, we also provide the raw measurements and invite the interested reader to experiment with both versions. Furthermore, a forward-modeling approach, as opposed to de-convolution, may yield a more accurate estimation of the contribution of the spot size to the measured IPRF.  

Lastly, this experiment was performed at a higher temperature than \textit{Kepler}'s operating temperature (-45$^{\circ}$ C vs -85$^{\circ}$ C). Based on our analytical models of the temperature dependence of silicon's absorption depth, we expect a wavelength shift of $\Delta\lambda\approx - 18$ nm of the measured response. In other words, the IPRF measured at 550 nm at -45$^{\circ}$ C, is more like the IPRF at 530 nm at -85$^{\circ}$ C. However, this difference does not appear to be large (see Appendix \ref{sec:tempratureWavelengthEffects}). It is not yet clear to us that the difference in the combined ``effective'' IPRF would be significant. Nevertheless, this is an important aspect of this experiment and is the subject of future work.

\section{Conclusions}
The spot scanning technique has been shown to be effective at measuring the IPRF of imaging sensors. Our measurements show that \textit{Kepler}'s IPRF is highly nonuniform and wavelength-dependent. Our new spot scanner uses modern technology (stable light sources, precise and reliable translation stages) to allow routine, non-destructive measurements of IPRF for a wide range of imaging sensors.

The calibrated IPRF maps are available to the public at our GitHub repository: \url{https://github.com/keplerIPRF/keplerIPRF}. 

\acknowledgments     
 
This research is funded by NASA Astrophysics Data Analysis Program award NNX16AF43G. Initial results from this work were presented at the SPIE Astronomical Telescopes and Instrumentation meeting in Austin, Texas in June, 2018, as paper number 106985J. The authors would like to thank Charlie Sobeck, the Kepler Project Manager at NASA Ames, for assistance with locating and securing a Kepler CCD for these tests. We appreciate John Troeltzsch, Anne Ayers, and Rick Ortiz at Ball Aerospace in Boulder for providing information regarding the packaging and electronics associated with the \textit{Kepler} CCD. We're grateful to Charles Slaughter and Gary Sims at Spectral Instruments in Arizona for modifying a camera to use the \textit{Kepler} CCD. We thank Kevin Fogarty of Canon USA for providing the 120MXS CMOS sensor and evaluation kit. Lastly, we thank Jonathan Hoover for significantly improving the stability of the data acquisition software.

\appendix
\section{Managing the Effects of Spot Size}
\label{sec:deconvolvingIPRF}
\begin{figure*}
    \centering
    \includegraphics[width=\textwidth]{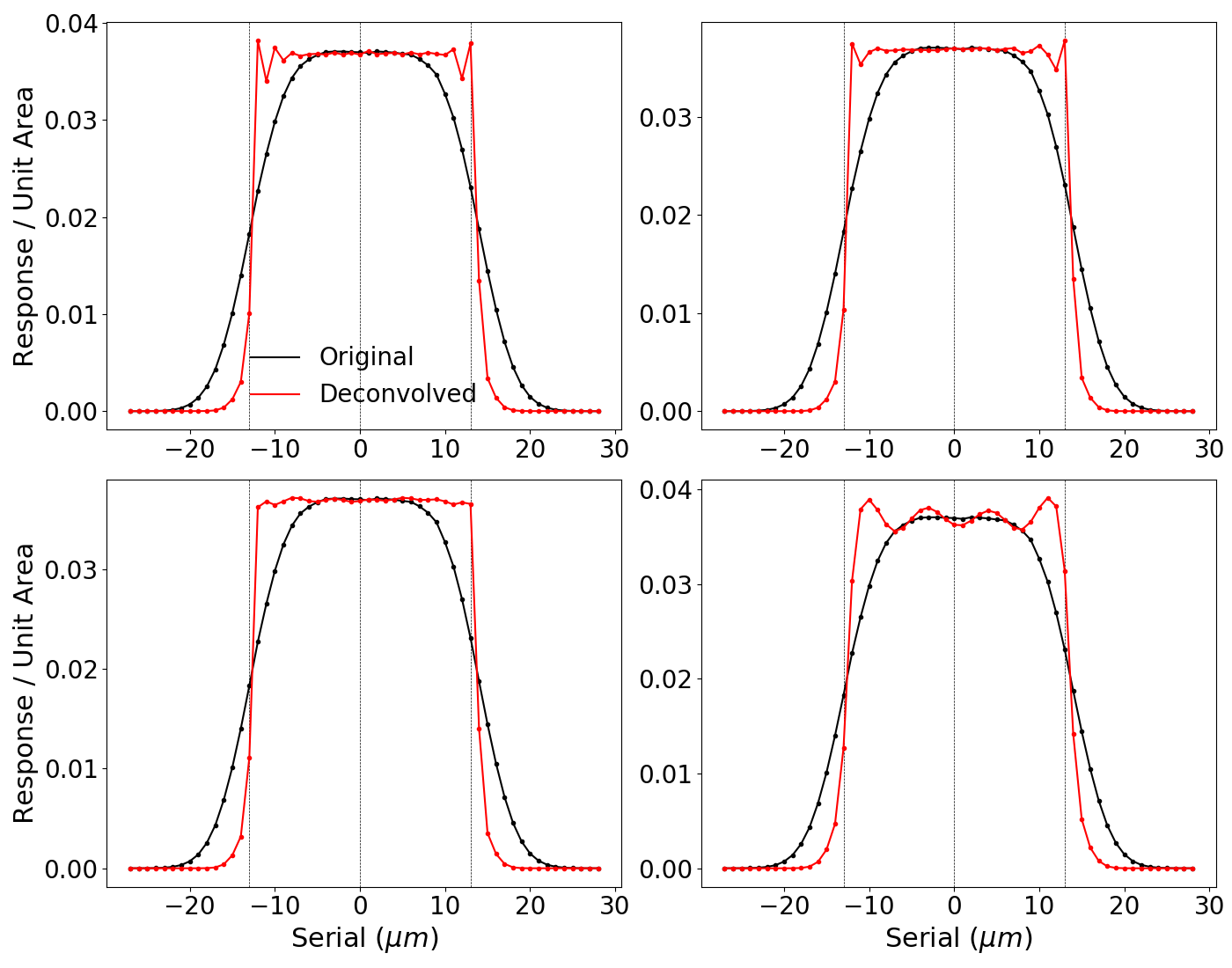}
    \caption{Simulations of IPRF measurements with spots of different sizes show that for a 27 $\mu m$ pixel, spots with FWHM $<$ 4 $\mu m$ introduce negligible or manageable uncertainty to the inferred IPRF. However, spots with FWHM $\gtrsim\frac{1}{4}$ of pixel size result in ambiguous measurements that cannot be easily calibrated/interpreted.}
    \label{fig:simAndDeconPlots}
\end{figure*}

The finite extent of the illuminating spot can give rise to ``blurring'' artifacts that mask/mimic the actual structure of the IPRF (See Section \ref{sec:spotSizeEffects}). If the spot size is sufficiently small, as compared to the pixel size, these effects can be separated from the true IPRF signal (Fig.  \ref{fig:simAndDeconPlots}. In this case (and likely in other back-illuminated CCDs), if the spot size is smaller than $\approx$ $\frac{1}{6}$ of the pixel size, the effects are either negligible or they can be mitigated through de-convolution (if the spot size is well characterized and the IPRF does not exhibit high frequency structure) or forward-modeling. However, when the spot FWHM becomes larger than $\approx$ $\frac{1}{4}$ of the pixel size, the effects of spot size and the intrinsic IPRF become difficult to isolate and calibrate (Fig.  \ref{fig:simAndDeconPlots}, 8 $\mu m$ scan). 

\begin{figure*}
    \centering
    \includegraphics[width=\textwidth]{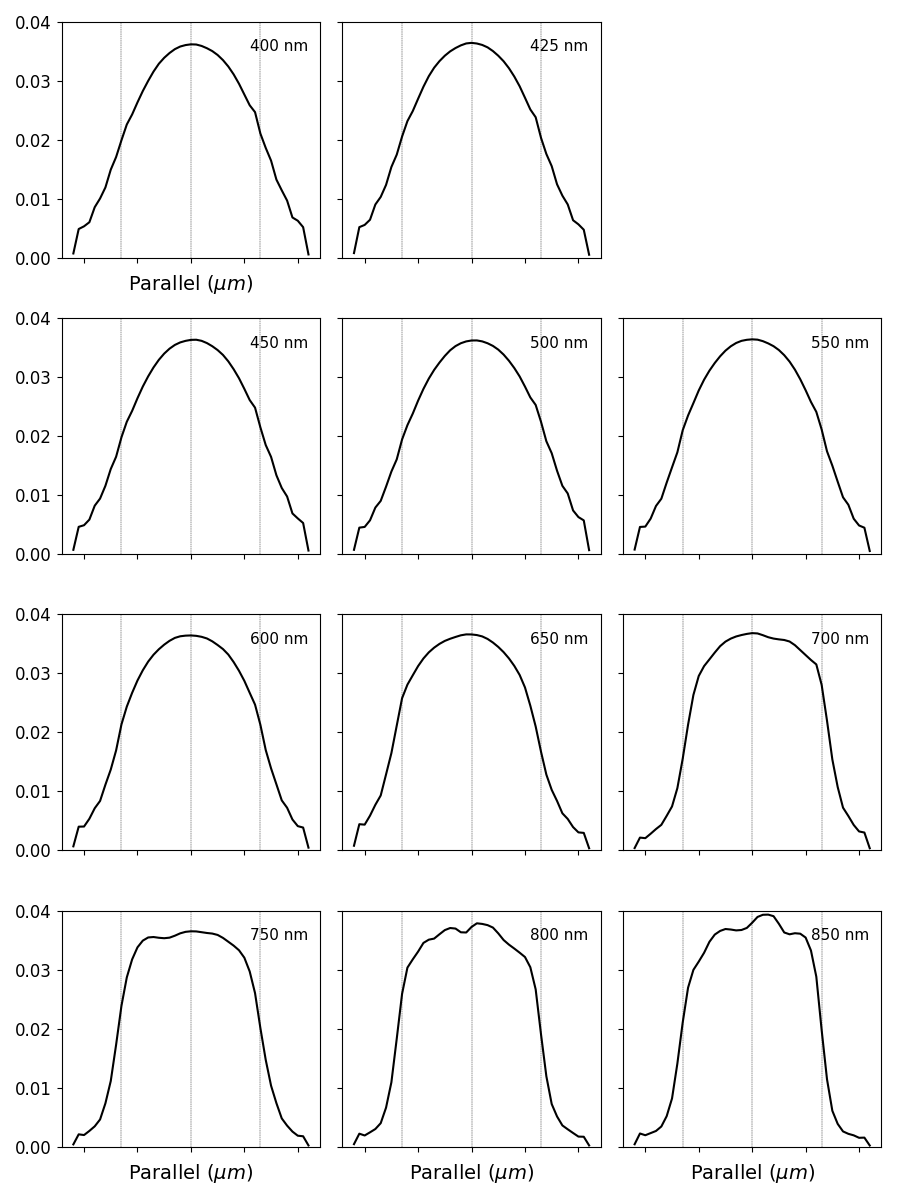}
    \caption{The original measured profiles (volume-normalized) are calibrated using Lucy-Richardson de-convolution, to account for the blurring due to the finite size of the spot.}
    \label{fig:raw_and_decon_iprfs}
\end{figure*}

A comparison of the raw measured IPRFs and the de-convolved profiles are shown in Fig.  \ref{fig:raw_and_decon_iprfs}. At each wavelength, we performed the de-convolution using the spot sizes estimated by assuming one of three IPRF models for the Canon 120MXS sensor (which was used to characterize the spot size): delta-like, perfectly uniform (rect) and Gaussian. The choice of IPRF here does not significantly affect the reconstruction, though we expect that the Gaussian assumption gives the best estimation of the spot size, because the 120MXS is a front-illuminated sensor with microlenses. The delta-like and uniform (rect) models are too extreme and simplistic to accurately model the PRF; however, they are useful as a simple means to place upper and lower bounds on this estimation. 

\clearpage
\section{Temperature-Wavelength Absorption Depth Changes in Silicon}
\label{sec:tempratureWavelengthEffects}

The measurements of the IPRF of the \textit{Kepler} CCD90 sensor were performed at -45$^{\circ}$ C, because our camera could not achieve the \textit{Kepler} operating temperature of -85$^{\circ}$ C. Because the absorption depth in silicon depends on temperature (as well as wavelength), the results we obtained cannot be directly applied to the data acquired on orbit. We investigated the effects of this temperature difference on the IPRF using analytic models of absorption depth and by comparing IPRF measurements made at -45$^{\circ}$ C and +10$^{\circ}$ C. In this appendix, we present our analysis of the temperature dependence of the IPRF and propose a method to calibrate our measurements to make them directly applicable on on-orbit data.

\subsection{Absorption Depth in Silicon}

\begin{figure}
    \centering
    \includegraphics[width=\linewidth]{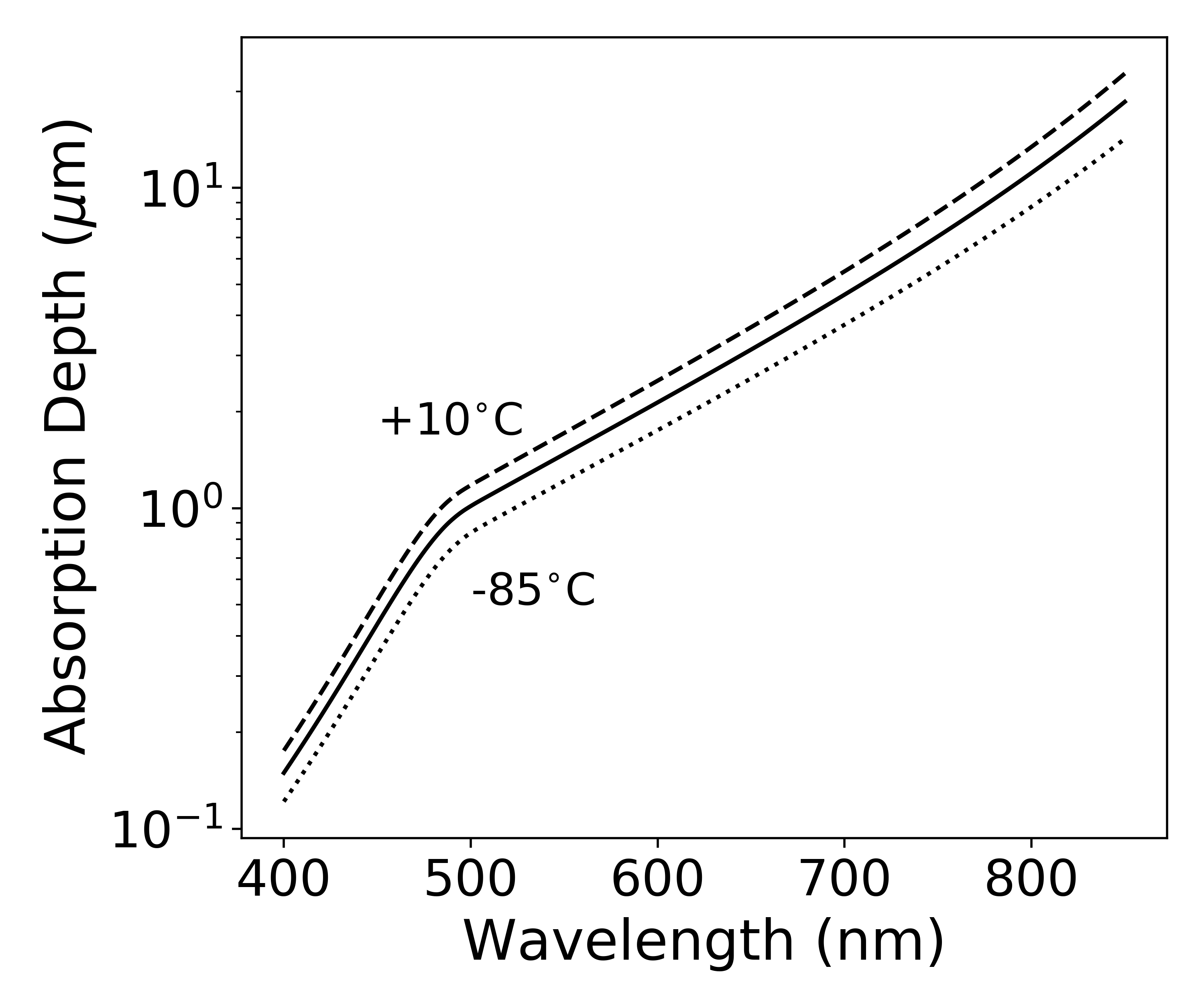}
    \caption{Absorption depth in silicon across the \textit{Kepler} passband, calculated for -85$^{\circ}$ C (dashed line), -45$^{\circ}$ C (solid line), and +10$^{\circ}$ C (dotted line) using the model of Rajkanan et al.\cite{Rajkanan1979}.}
    \label{fig:Absorption_Depth_in_Si}
\end{figure}

We calculate the absorption depth in silicon, as a function of wavelength and temperature, using the model developed by Rajnakan et al.\cite{Rajkanan1979}. This model applies over the temperature range of 20 - 500 K and photon energies of 1.1 - 4 eV. The calculated absorption depth for silicon temperatures at -85$^{\circ}$ C (\textit{Kepler} operating temperature), -45$^{\circ}$ C (spot scan temperature) and +10$^{\circ}$ C (elevated temperature spot scan) are given in Fig.  \ref{fig:Absorption_Depth_in_Si}.

\begin{figure}
    \centering
    \includegraphics[width=\linewidth]{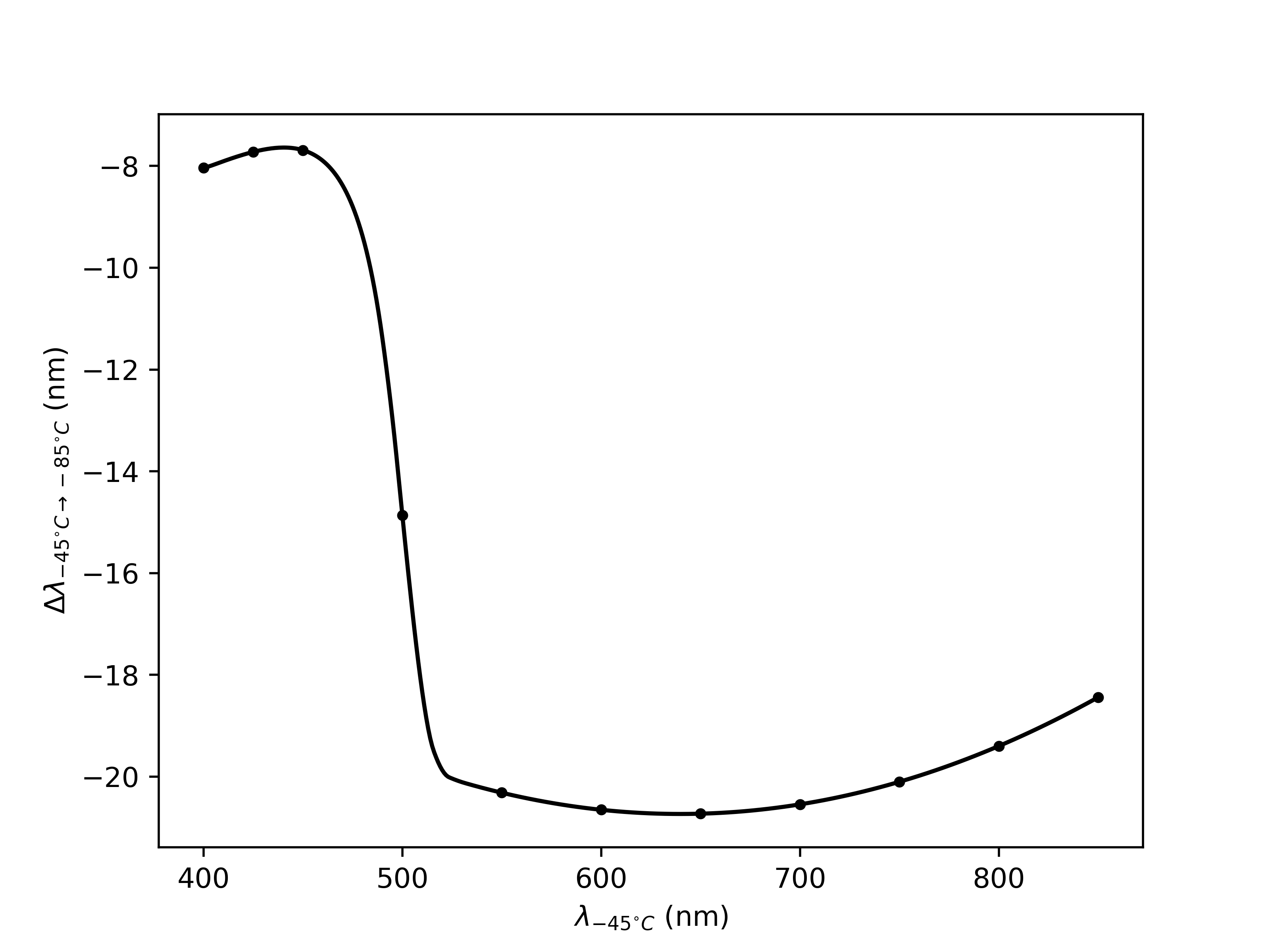}
    \caption{As the absorption depth in silicon increases with decreasing temperature, the equivalent-depth wavelength is shifted blue-wards. Here we plot the $\Delta\lambda$ across the \textit{Kepler} passband, due to a temperature change from -45$^{\circ}$ C to -85$^{\circ}$ C.}
    \label{fig:wavelengthShift}
\end{figure}

Overall, the absorption depth at a particular wavelength increases as the temperature decreases. Using the calculated absorption depths (Fig.  \ref{fig:Absorption_Depth_in_Si}), we determine the wavelength with equivalent absorption depth, when the temperature is changed from -45$^{\circ}$ C to -85$^{\circ}$ C. This wavelength shift is shown in Fig.  \ref{fig:wavelengthShift}. 

\begin{figure}
    \centering
    \includegraphics[width=\linewidth]{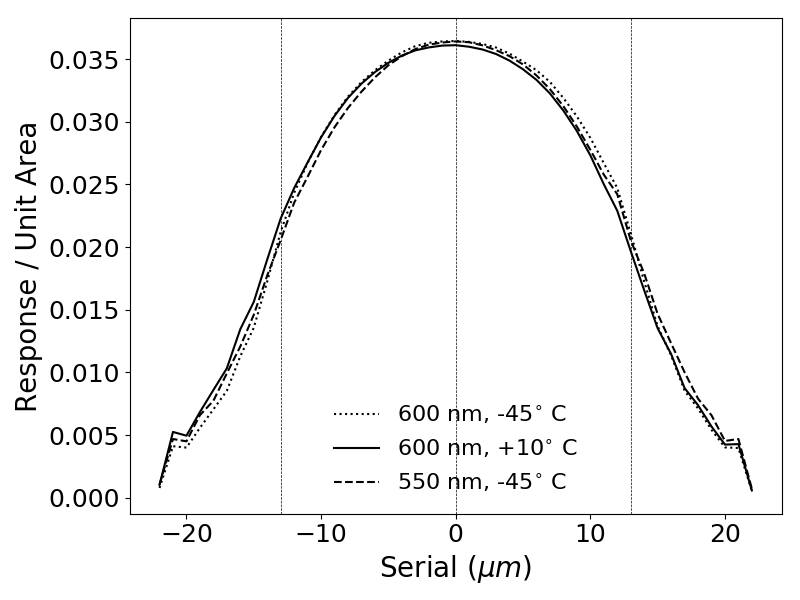}
    \caption{The IPRF measured at 600 nm at two different temperatures (-45$^{\circ}$ C and +10$^{\circ}$ C), compared with the IPRF at 550 nm (-45$^{\circ}$ C).}
    \label{fig:temperatureComparison}
\end{figure}

The largest wavelength shifts occur near 650 nm, with a $\Delta\lambda$ of 20 nm. Practically, this means the IPRF measured at 650 nm at -45$^{\circ}$ C, represents the response at 630 nm at -85$^{\circ}$ C. Although the IPRF we measured has a strong wavelength dependence across the entire \textit{Kepler} passband (400 - 850 nm), the incremental change over 50 nm (the smallest interval we measured) isn't large. We compare the IPRF acquired at two different temperatures (-45$^{\circ}$ C and +10$^{\circ}$ C) at 600 nm, to the IPRF measured at 550 nm (-45$^{\circ}$ C) in Fig.  \ref{fig:temperatureComparison}.

Based on our analysis, the differences between the IPRF measured in the lab at -45$^{\circ}$ C and the actual IPRF on orbit (at -85$^{\circ}$ C) appear minor. The median wavelength shift across the \textit{Kepler} passband is $\sim$18 nm, which is comparable to the spectral bandwidth of the illuminating spot ($\Delta\lambda\approx15$ nm) and much smaller than the wavelength interval between measurements (50 nm). Therefore, we expect minimal changes to the final \textit{Kepler} IPRF, when all individual measurements (450 nm, 500 nm, \textit{etc}.) are combined into a single broadband response function (see Section \ref{sec:kepler_iprf}). Nevertheless, this is an important aspect of the IPRF characterization (as it was performed) and the final significance of this effect can only be determined once these IPRF measurements are incorporated into a data reduction pipeline.


\bibliography{report}   

\begin{thebibliography}{10}

\bibitem{Borucki2010}
W.~J. {Borucki}, D.~{Koch}, G.~{Basri}, N.~{Batalha}, T.~{Brown},
  D.~{Caldwell}, J.~{Caldwell}, J.~{Christensen-Dalsgaard}, W.~D. {Cochran},
  E.~{DeVore}, E.~W. {Dunham}, A.~K. {Dupree}, T.~N. {Gautier}, J.~C. {Geary},
  R.~{Gilliland}, A.~{Gould}, S.~B. {Howell}, J.~M. {Jenkins}, Y.~{Kondo},
  D.~W. {Latham}, G.~W. {Marcy}, S.~{Meibom}, H.~{Kjeldsen}, J.~J. {Lissauer},
  D.~G. {Monet}, D.~{Morrison}, D.~{Sasselov}, J.~{Tarter}, A.~{Boss},
  D.~{Brownlee}, T.~{Owen}, D.~{Buzasi}, D.~{Charbonneau}, L.~{Doyle},
  J.~{Fortney}, E.~B. {Ford}, M.~J. {Holman}, S.~{Seager}, J.~H. {Steffen},
  W.~F. {Welsh}, J.~{Rowe}, H.~{Anderson}, L.~{Buchhave}, D.~{Ciardi},
  L.~{Walkowicz}, W.~{Sherry}, E.~{Horch}, H.~{Isaacson}, M.~E. {Everett},
  D.~{Fischer}, G.~{Torres}, J.~A. {Johnson}, M.~{Endl}, P.~{MacQueen}, S.~T.
  {Bryson}, J.~{Dotson}, M.~{Haas}, J.~{Kolodziejczak}, J.~{Van Cleve},
  H.~{Chandrasekaran}, J.~D. {Twicken}, E.~V. {Quintana}, B.~D. {Clarke},
  C.~{Allen}, J.~{Li}, H.~{Wu}, P.~{Tenenbaum}, E.~{Verner}, F.~{Bruhweiler},
  J.~{Barnes}, and A.~{Prsa}, ``{Kepler Planet-Detection Mission: Introduction
  and First Results},'' {\em Science}~{\bf 327}, p.~977, Feb. 2010.

\bibitem{Bryson2010}
S.~T. {Bryson}, P.~{Tenenbaum}, J.~M. {Jenkins}, H.~{Chandrasekaran},
  T.~{Klaus}, D.~A. {Caldwell}, R.~L. {Gilliland}, M.~R. {Haas}, J.~L.
  {Dotson}, D.~G. {Koch}, and W.~J. {Borucki}, ``{The Kepler Pixel Response
  Function},'' {\em Ap. J. Lett.}~{\bf 713}, pp.~L97--L102, Apr. 2010.

\bibitem{Morris2014}
R.~L. {Morris}, S.~{Bryson}, J.~M. {Jenkins}, and J.~C. {Smith}, ``{Improving
  Kepler Pipeline Sensitivity with Pixel Response Function Photometry.},'' in
  {\em American Astronomical Society Meeting Abstracts \#224},  {\em American
  Astronomical Society Meeting Abstracts} {\bf 224}, p.~120.08, June 2014.

\bibitem{Christiansen2012}
J.~L. {Christiansen}, J.~M. {Jenkins}, D.~A. {Caldwell}, C.~J. {Burke},
  P.~{Tenenbaum}, S.~{Seader}, S.~E. {Thompson}, T.~S. {Barclay}, B.~D.
  {Clarke}, J.~{Li}, J.~C. {Smith}, M.~C. {Stumpe}, J.~D. {Twicken}, and
  J.~{Van Cleve}, ``{The Derivation, Properties, and Value of Kepler's Combined
  Differential Photometric Precision},'' {\em PASP}~{\bf 124}, p.~1279, Dec.
  2012.

\bibitem{Monet2010}
D.~G. {Monet}, J.~M. {Jenkins}, E.~W. {Dunham}, S.~T. {Bryson}, R.~L.
  {Gilliland}, D.~W. {Latham}, W.~J. {Borucki}, and D.~G. {Koch},
  ``{Preliminary Astrometric Results from Kepler},'' {\em arXiv e-prints} ,
  p.~arXiv:1001.0305, Jan 2010.

\bibitem{Vanderburg2014}
A.~{Vanderburg} and J.~A. {Johnson}, ``{A Technique for Extracting Highly
  Precise Photometry for the Two-Wheeled Kepler Mission},'' {\em PASP}~{\bf
  126}, p.~948, Oct. 2014.

\bibitem{Luger2016}
R.~{Luger}, E.~{Agol}, E.~{Kruse}, R.~{Barnes}, A.~{Becker},
  D.~{Foreman-Mackey}, and D.~{Deming}, ``{EVEREST: Pixel Level Decorrelation
  of K2 Light Curves},'' {\em Astronomical Journal}~{\bf 152}, p.~100, Oct.
  2016.

\bibitem{Luger2018}
R.~{Luger}, E.~{Kruse}, D.~{Foreman-Mackey}, E.~{Agol}, and N.~{Saunders},
  ``{An Update to the EVEREST K2 Pipeline: Short Cadence, Saturated Stars, and
  Kepler-like Photometry Down to Kp=15},'' {\em Astronomical Journal}~{\bf
  156}, p.~99, Sept. 2018.

\bibitem{Lauer1999}
T.~R. {Lauer}, ``{The Photometry of Undersampled Point-Spread Functions},''
  {\em PASP}~{\bf 111}, pp.~1434--1443, Nov. 1999.

\bibitem{Koch2000}
D.~G. Koch, W.~J. Borucki, E.~W. Dunham, W.~L. Jenkins, J.~M., and F.~C.
  Witteborn, ``Ccd photometry tests for a mission to detect earth-sized planets
  in the extended solar neighborhood,'' 2000.

\bibitem{Jorden1994}
P.~R. Jorden, J.-M. Deltorn, and A.~P. Oates, ``Nonuniformity of ccds and the
  effects of spatial undersampling,'' {\em Proc.SPIE}~{\bf 2198}, pp.~2198 --
  2198 -- 15, 1994.

\bibitem{Piterman2002}
A.~{Piterman} and Z.~{Ninkov}, ``{Subpixel sensitivity maps for a
  back-illuminated charge-coupled device and the effects of nonuniform response
  on measurement},'' {\em Optical Engineering}~{\bf 41}, June 2002.

\bibitem{Kavaldjiev1998}
D.~Kavaldjiev and Z.~Ninkov, ``Subpixel sensitivity map for a charge-coupled
  device,'' {\em Optical Engineering}~{\bf 37}, pp.~37 -- 37 -- 7, 1998.

\bibitem{Vorobiev2018}
D.~Vorobiev, Z.~Ninkov, D.~Caldwell, and S.~Mochnacki, ``Direct measurement of
  the intra-pixel response function of the kepler space telescope's ccds,''
  {\em Proc. SPIE}~{\bf 10698}, pp.~10698 -- 10698 -- 14, 2018.

\bibitem{vancleve2016}
J.~E. {Van Cleve} and D.~A. {Caldwell}, ``{Kepler Instrument Handbook},'' tech.
  rep., Apr. 2016.

\bibitem{Rajkanan1979}
K.~{Rajkanan}, R.~{Singh}, and J.~{Shewchun}, ``{Absorption coefficient of
  silicon for solar cell calculations},'' {\em Solid State Electronics}~{\bf
  22}, pp.~793--795, Sept. 1979.

\end{thebibliography}
\bibliographystyle{spiebib}   
\end{spacing}
\begin{doublespace}
\end{doublespace}
\end{document}